\documentclass[Afour,sageh,times]{sagej}

\usepackage{moreverb}
\usepackage{amsmath,amsthm}
\usepackage{amssymb}
\usepackage{mathrsfs}
\usepackage{bbm}
\usepackage{mathtools}
\usepackage{bm}
\usepackage{url}
\usepackage{amsfonts}
\usepackage{cases}
\usepackage{graphicx}
\usepackage{float}
\usepackage{multirow}
\usepackage{color}
\usepackage{algorithm}
\usepackage{algpseudocode}
\usepackage{graphicx}
\usepackage{subfigure}
\usepackage{epstopdf}
\usepackage{lipsum}
\usepackage{tikz}
\usetikzlibrary{positioning,shapes}
\usepackage{hyperref}
\hypersetup{
    colorlinks=true,
    linkcolor=black,
    citecolor=black,
    urlcolor=blue,
}

\graphicspath{{./}{Fig/}}

\DeclareMathOperator*{\argmax}{argmax}

\allowdisplaybreaks[4]

\algnewcommand\algorithmicswitch{\textbf{switch}}
\algnewcommand\algorithmiccase{\textbf{case}}
\algnewcommand\algorithmicassert{\texttt{assert}}
\algnewcommand\Assert[1]{\State \algorithmicassert(#1)}%

\algdef{SE}[SWITCH]{Switch}{EndSwitch}[1]{\algorithmicswitch\ #1\ \algorithmicdo}{\algorithmicend\ \algorithmicswitch}%
\algdef{SE}[CASE]{Case}{EndCase}[1]{\algorithmiccase\ #1}{\algorithmicend\ \algorithmiccase}%
\algtext*{EndSwitch}%
\algtext*{EndCase}%

\newtheorem{definition}{Definition}[section]

\newtheorem{proposition}{Proposition}[section]
\newtheorem{theorem}{Theorem}[section]
\newtheorem{remark}{Remark}[section]
\newtheorem{example}{Example}[section]

\newtheorem{problem}{Problem}[section]

\newcommand\BibTeX{{\rmfamily B\kern-.05em \textsc{i\kern-.025em b}\kern-.08em
T\kern-.1667em\lower.7ex\hbox{E}\kern-.125emX}}

\def\volumeyear{2023}
\begin{document}

\setcitestyle{aysep={,}}

\title{Online Control Synthesis for Uncertain Systems under Signal Temporal Logic Specifications}

\runninghead{Yu et al.}

\author{Pian Yu\affilnum{1}, Yulong Gao\affilnum{1}, Frank J. Jiang\affilnum{2,3}, Karl H. Johansson\affilnum{2,3}, and Dimos V. Dimarogonas\affilnum{2,3}}

\affiliation{\affilnum{1}Department of Computer Science, University of Oxford, UK\\
\affilnum{2}Division of Decision and Control
Systems, KTH Royal Institute of Technology, Stockholm, Sweden\\   \affilnum{3}Digital Futures,  Stockholm, Sweden}

\corrauth{Pian Yu, Department of Computer Science, University of Oxford, UK.}

\email{pian.yu@cs.ox.ac.uk}

\begin{abstract}
This paper studies the online control synthesis problem for uncertain discrete-time systems subject to signal temporal logic (STL) specifications. Different from existing techniques, this work proposes an approach based on STL, reachability analysis, and temporal logic trees. Firstly, a real-time version of STL semantics and a tube-based temporal logic tree (tTLT) are proposed. We show that the tTLT is an underapproximation for the STL formula, in the sense that a trajectory satisfying an tTLT also satisfies the corresponding STL formula.  Secondly, an online control synthesis algorithm is designed. It is shown that when the STL formula is robustly satisfiable and the initial state of the system belongs to the initial root node of the tTLT, it is guaranteed that the trajectory generated by the control synthesis algorithm satisfies the STL formula. The effectiveness of the proposed approach is verified by a simulation example and a practical experiment.
\end{abstract}

\keywords{Signal temporal logic, uncertain systems, online control synthesis, tube-based temporal logic tree, and reachability analysis}

\maketitle

\section{Introduction}
\subsection{Motivation and Related Work}
Rapid growth of robotic applications, such as autonomous vehicles and service robots, has stimulated the need of new control synthesis approaches to safely accomplish  more complex objectives such as nondeterministic, periodic, or sequential tasks. Temporal logics, such as linear temporal logic (LTL) \citep{Baier2008}, metric interval temporal logic (MITL) \citep{koymans1990}, and signal temporal logic (STL) \citep{maler2004}, have shown capability in expressing such objectives for dynamical systems in the last decade. Various control approaches have been developed accordingly.

LTL focuses on the Boolean satisfaction of properties by given signals while MITL is a continuous-time extension that allows to express temporal constraints. Existing control approaches that use LTL or MITL mainly rely on a finite abstraction of the system dynamics and a language equivalent automata \citep{gastin2001} or timed-automata \citep{alur1996} representation of the LTL or MITL specification. The controller is synthesized by solving a game over the product automata \citep{belta2007,belta2017,zhou2016}. Other control approaches include optimization-based \citep{wolff2016,fu2015} and sampling-based methods \citep{vasile2013,kantaros2018}. STL is a more recently developed temporal logic, which allows the specification of properties over dense-time. Due to a number of advantages, such as explicitly treating real-valued signals \citep{maler2004}, and admitting qualitative semantics  \citep{fainekos2009}, control synthesis under STL specifications has gained popularity in the last few years.

Different from LTL or MITL, automata-based methods have not been developed for STL specifications to the same extent due to their complexity. Existing approaches that deal with control synthesis under STL specifications include optimization \citep{raman2015,raman2014,sadraddini2015} and barrier function methods \citep{lindemann2018,lindemann2019,yang2020continuous}. Optimization methods are mainly used for discrete-time systems. The idea is to encode STL formulas as mixed-integer constraints, and then the satisfying controller can be obtained by solving a series of optimization problems \citep{raman2015,raman2014}. An extension of the mixed-integer formulation is investigated for linear systems with additive bounded disturbances in \citet{sadraddini2015}, where the controller is obtained by solving the optimization problem at each time step in a receding horizon fashion. One drawback of this approach is the exponential computational complexity which makes it difficult to be applied to STL formulas with long time
horizons. Barrier function methods are mainly used for continuous-time systems. The idea is to transfer the STL formula into one or several (time-varying) control barrier functions, and then obtain feedback control laws by solving quadratic programs \citep{lindemann2018,lindemann2019}. This method is computationally efficient. However, as the existence and design of barrier functions are still open problems, it currently mainly applies to deterministic affine systems. In \citet{yang2020continuous}, the authors consider linear cyber-physical systems with continuous-time dynamics and discrete-time controllers. The proposed offline trajectory planner is based on a mixed integer quadratic programming that utilizes control barrier functions to generate satisfying trajectories in continuous-time.
Other control synthesis approaches include sampling-based \citep{vasile2017,karlsson2020sampling} and learning-based methods \citep{venkataraman2020tractable,kapoor2020model}. In addition, control synthesis for multi-agent systems and STL specifications is recently considered in \citet{lindemann2019feedback,buyukkocak2021planning,sun2022multi}.

We note that although various methods exist for the control synthesis under STL specifications, guaranteeing robustness against uncertainties is still a challenging problem. The core contribution of this paper is on robust control synthesis for uncertain systems under STL specifications.

\subsection{Main Contributions and Organization}
Motivated by the above considerations, this work considers the online control synthesis problem for uncertain discrete-time systems under STL specifications. The paper is inspired by \citet{chen2018}, where relationships between primitive STL formulas and reachable sets are developed, and \citet{gao2020}, where the notion of temporal logic tree is proposed for LTL. However, we note that it is far from straightforward to extend these results to general STL formulas. The contributions of our paper are summarized as follows:
\begin{itemize}
	\item [(i)] A real-time version of satisfaction relation and a tube-based temporal logic tree (tTLT) are proposed for STL formulas. A correspondence between STL formulas and tTLT is established via reachability analysis on the underlying systems. An algorithm is proposed for the automated construction of tTLT. Note that the tTLTs in this paper are different from the TLTs defined for LTL formulas in \citet{gao2020}, due to the time constraints encoded in the STL formulas.
	\item [(ii)] We show that the tTLT is an underapproximation for a broad fragment of STL formulas, \textit{i.e.,} all the trajectories that satisfy the tTLT also satisfy the corresponding STL formula. 
	\item [(iii)] We propose an online control synthesis algorithm based on the constructed tTLT from the STL formula. When the STL formula is robustly satisfiable and the initial state of the system belongs to the initial root node of the tTLT, it is proven that the trajectory generated by the proposed online control synthesis algorithm satisfies the STL formula.
\end{itemize}

The remainder of the paper is organized as follows. In Section
2, preliminaries and the problem under consideration are formulated. In Section 3, definitions of real-time STL semantics and tTLT are introduced. Section 4 establishes a semantic connection between STL and tTLT. Section 5 deals with the online control synthesis problem. The results are validated by simulations and experiments in Section 6. Conclusions are given in Section~7.

\textbf{Notation.} Let $\mathbb{R}:=(-\infty, \infty)$, $\mathbb{R}_{\ge 0}:=[0, \infty)$. Let $\mathbb{N}$ be the set of natural numbers. Denote $\mathbb{R}^n$ as the $n$ dimensional real vector space, $\mathbb{R}^{n\times m}$ as the $n\times m$ real matrix space. Given a vector $x\in \mathbb{R}^n$, define $\|x\|$ and $x^T$ as the Euclidean norm and the transpose of vector $x$, respectively. Given a set $\Omega$, $\overline{\Omega}$ denotes its complement, $2^{\Omega}$ denotes its powerset, and $|\Omega|$ denotes its cardinality. The operators $\cup$ and $\cap$ represent set union and set intersection, respectively. In addition, we use $\wedge$ to denote the logical operator AND and $\vee$ to denote the logical operator OR.  The set difference $A\setminus B$ is defined by $A\setminus B:=\{x: x\in A \;\wedge\; x\notin B\}$.

\section{Preliminaries and Problem Formulation}\label{Sec:Preliminaries}

\subsection{Systems dynamics}\label{Plantmodel}

Consider an uncertain discrete-time control system of the form
\begin{equation}\label{x0}
	\begin{aligned}
		x_{k+1}&=f(x_k, u_k, w_k),\\
	\end{aligned}
\end{equation}
where $x_k:=x(t_k)\in \mathbb{R}^n, u_k:=u(t_k)\in U,  w_k:=w(t_k)\in W, k\in \mathbb{N}$ are the state, control input, and disturbance at time $t_k$, respectively. The time sequence $\{t_k\}$ can be seen as a sequence of sampling instants, which satisfy $0=t_0<t_1<\cdots$. The control input is constrained to a compact set $U\subset \mathbb{R}^m$ and the disturbance is constrained to a compact set $W\subset \mathbb{R}^l$. In the following, let us define the control policy.

\begin{definition}\label{controlpolicy}
A \emph{control policy} $\bm{\nu}=\nu_0\nu_1\ldots \nu_k\ldots$ is a sequence of maps $\nu_k: \mathbb{R}^n\rightarrow U$, $\forall k\in \mathbb{N}$. Denote by $\mathcal{U}_{\ge k}$ the set of all control policies that start from time $t_k$.
\end{definition}

One can see from Definition \ref{controlpolicy} that a control policy is a sequence of time-dependent functions, each of which maps from the state space to the input space.

\begin{definition}
	A disturbance signal $\bm{w}=w_0w_1\ldots w_k\ldots$ is called \emph{admissible} if $w_k\in W, \forall k\in \mathbb{N}$. Denote by $\mathcal{W}_{\ge k}$ the set of all admissible disturbance signals that start from time $t_k$.
\end{definition}

The solution of (\ref{x0}) is defined as a discrete-time signal $\bm{x}:=x_0x_1\ldots$. We call $\bm{x}$ a \emph{trajectory} of (\ref{x0}) if there exists a control policy $\bm{\nu}\in \mathcal{U}_{\ge 0}$ and a disturbance signal $\bm{w}\in \mathcal{W}_{\ge 0}$ satisfying (\ref{x0}), \textit{i.e.,}
\begin{equation*}
	x_{k+1}=f(x_k, \nu_k(x_k), w_k),  \forall k\in \mathbb{N}.
\end{equation*}
We use ${\bm{x}}_{x_0}^{\bm{\nu}, \bm{w}}(t_k)$ to denote the trajectory point reached at time $t_k$ under the control policy $\bm{\nu}$ and the disturbance $\bm{w}$ from state $x_0$ at time $t_0$.

The deterministic system is defined by
\begin{equation}\label{x1}
	x_{k+1}=f_d(x_k, u_k)
\end{equation}
and ${\bm{x}}_{x_0}^{\bm{\nu}}(t_k)$ denotes the solution at time $t_k$ of the deterministic system when the control policy is $\bm{\nu}$ and the initial state is
$x_0$ at time $t_0$.

\subsection{Signal temporal logic}

We use STL to concisely specify the desired system behavior. STL \citep{maler2004} is a predicate logic consisting of predicates $\mu$, which are defined through a predicate function $g_\mu: \mathbb{R}^n\to \mathbb{R}$ as
\begin{equation*}
	\mu:=\left\{\begin{aligned}
		\top, & \quad \text{if } \quad g_\mu(x)\ge 0 \\
		\bot, & \quad \text{if } \quad g_\mu(x)<0.
	\end{aligned}\right.
\end{equation*}

The syntax of STL is given by
\begin{eqnarray}\label{STLdef}
	\varphi ::= \top \mid \mu \mid \neg  \varphi \mid \varphi_1 \wedge \varphi_2 \mid  \varphi_1 \mathsf{U}_{\text{I}} \varphi_2,
\end{eqnarray}
where $\varphi, \varphi_1, \varphi_2$ are STL formulas and ${\text{I}}$ is a closed interval of $\mathbb{R}$ of the form $[a, b]$ with $a, b\in \mathbb{R}_{\ge 0}\cup \infty$ and $a\le b$. 

The validity of an STL formula $\varphi$ with respect to a discrete-time
signal $\bm{x}$ at time $t_k$, is defined inductively as follows \citep{raman2015}:
\begin{eqnarray*}
	(\bm{x}, t_k) \vDash \mu &\Leftrightarrow& g_\mu(\bm{x}(t_k))\ge 0, \\
	(\bm{x}, t_k) \vDash  \neg  \varphi &\Leftrightarrow& \neg((\bm{x}, t_k) \vDash \varphi), \\
	(\bm{x}, t_k) \vDash \varphi_1 \wedge \varphi_2 &\Leftrightarrow& (\bm{x}, t_k) \vDash \varphi_1 \wedge  (\bm{x}, t_k) \vDash \varphi_2, \\
	(\bm{x}, t_k) \vDash \varphi_1 \mathsf{U}_{[a, b]} \varphi_2 &\Leftrightarrow& \exists t_{k'}\in [t_k+a, t_k+b]  \ \text{s.t.} \\
	& & (\bm{x}, t_{k'})  \vDash \varphi_2 \wedge \forall t_{k''}\in [t_k, t_{k'}], \\
	& & (\bm{x}, t_{k''}) \vDash \varphi_1.
\end{eqnarray*}

The signal $\bm{x}=x_0x_1\ldots$ satisfies $\varphi$, denoted by $\bm{x} \vDash \varphi$ if $(\bm{x}, t_0) \vDash \varphi$. By using the ``negation" operator $\neg$ and the ``conjunction" operator $\wedge$, we can define ``disjunction" $\varphi_1\vee \varphi_2=\neg (\neg \varphi_1 \wedge \neg \varphi_2)$. And by employing the until operator $\mathsf{U}_{\text{I}}$, we can define ``eventually" $\mathsf{F}_{\text{I}} \varphi=\top \mathsf{U}_{\text{I}} \varphi$ and ``always" $\mathsf{G}_{\text{I}} \varphi=\neg \mathsf{F}_{\text{I}} \neg\varphi$.

\begin{definition}\citep{dokhanchi2014}
	The  time horizon $\|\varphi\|$ of an STL formula $\varphi$ is inductively defined as
	\begin{equation*}
		\|\varphi\|=\begin{cases}
			0, & \mbox{if } \varphi=\mu \\
			\|\varphi_1\|, & \mbox{if } \varphi=\neg\varphi_1 \\
			\max\{\|\varphi_1\|, \|\varphi_2\|\}, & \mbox{if } \varphi= \varphi_1\wedge\varphi_2 \\
			b+\max\{\|\varphi_1\|, \|\varphi_2\|\}, & \mbox{if } \varphi= \varphi_1\mathsf{U}_{[a, b]}\varphi_2.
		\end{cases}
	\end{equation*}
\end{definition}

\begin{definition}(Satisfiability)\label{Def:satisfibility}
	Consider the deterministic system (\ref{x1}) and the STL formula $\varphi$. We say $\varphi$ is satisfiable from the initial state $x_0$ if there exists a control policy $\bm{\nu}$ such that
	\begin{equation*}
		{\bm{x}}_{x_0}^{\bm{\nu}}  \vDash \varphi.
	\end{equation*}
\end{definition}

\begin{definition}(Robust satisfiability)\label{Def:feasibility}
	Consider the uncertain system (\ref{x0}) and the STL formula $\varphi$. We say $\varphi$ is robustly satisfiable from the initial state $x_0$ if there exists a control policy $\bm{\nu}$ such that
	\begin{equation*}
		{\bm{x}}_{x_0}^{\bm{\nu}, \bm{w}}  \vDash \varphi, \forall \bm{w}\in \mathcal{W}_{\ge 0}.
	\end{equation*}
\end{definition}

Given an STL formula $\varphi$, let
\begin{equation}\label{initialsatisfiableset}
	\mathbb{S}_{\varphi}:=\{x_0\in \mathbb{R}^n| \text{$\varphi$ is (robustly) satisfiable from $x_0$}\}
\end{equation}
denote the set of initial states from which $\varphi$ is (robustly) satisfiable.

\subsection{Reachability operators}

In this section, we define two reachability operators. The natural connection between reachability and temporal operators plays an important role to the approach proposed in this paper. The definitions of maximal and minimal reachable tube are given as follows.

\begin{definition}\label{Def:maxreachset}
	Consider the system (\ref{x0}), three sets $\Omega_1, \Omega_2, \mathcal{C} \subseteq \mathbb{R}^n$, and a time interval $[a, b]$. The maximal reachable tube from $\Omega_1$ to $\Omega_2$ is defined as
	\begin{eqnarray*}
		&&\mathcal{R}^M(\Omega_1,\Omega_2,\mathcal{C}, [a, b], k)\\
		&&\hspace{-0.5cm}=
		\left\{x_k\in \Omega_1 \;\middle\vert\; 
		\begin{array}{@{}l@{}}
			\exists \bm{\nu}\in \mathcal{U}_{\ge k},\forall \bm{w}\in
			\mathcal{W}_{\ge k},\; \text{s.t.}\;\\
			\exists t_{k'}\in [\max\{a, t_k\}, b], {\bm{x}}_{x_k}^{\bm{\nu},\bm{w}}(t_{k'})\in \Omega_2, \\
			\forall t_{k''}\in [t_k, t_{k'}], {\bm{x}}_{x_k}^{\bm{\nu}, \bm{w}}(t_{k''})\in \mathcal{C} 
		\end{array}
		\right\}\\
		&&\text{and} \ 
		t_{k}\in [0, b].
	\end{eqnarray*}
\end{definition}
The set $\mathcal{R}^M(\Omega_1,\Omega_2,\mathcal{C}, [a, b], k)$ collects all states in $\Omega_1$ at time $t_k$ from which there exists a control policy $\bm{\nu}\in \mathcal{U}_{\ge k}$ that, despite
the worst disturbance signals, drives the system to the target set $\Omega_2$ at some time instant $t_{k'}\in [\max\{a, t_k\}, b]$ while satisfying constraints defined by $\mathcal{C}$ prior to reaching the target.

\begin{definition}\label{Def:minreachset}
	Consider the system (\ref{x0}), two sets $\Omega_1, \Omega_2 \subseteq \mathbb{R}^n$, and a time interval $[a, b]$. The minimal reachable tube from $\Omega_1$ to $\Omega_2$ is defined as
	\begin{eqnarray*}
		&&	\mathcal{R}^m(\Omega_1,\Omega_2,[a, b], k) \\
		&&\hspace{-0.5cm}=\left\{x_k\in \Omega_1  \;\middle\vert\; 
		\begin{array} {@{}l@{}}
			\forall \bm{\nu}\in \mathcal{U}_{\ge k}, \exists \bm{w}\in \mathcal{W}_{\ge k},\; 	\text{s.t.}\; \\
			\exists t_{k'}\in [\max\{a, t_k\}, b], {\bm{x}}_{x_k}^{\bm{\nu},\bm{w}}(t_{k'})\in \Omega_2 
		\end{array}\right\},  \\
		&&\text{and} \ 
		t_{k}\in [0, b].
	\end{eqnarray*}
\end{definition}
The set $\mathcal{R}^m(\Omega_1,\Omega_2,[a, b], k)$ collects all states in $\Omega_1$ at time $t_k$ from which no matter what control policy $\bm{\nu}$ is applied, there exists a disturbance signal that drives the system to the target set $\Omega_2$ at some time instant $t_{k'}\in [\max\{a, t_k\}, b]$. In this definition, the constraint set $\mathcal{C}$ is redundant.

\subsection{Problem formulation}

Consider the following fragment of STL formulas, which is inductively defined as
	\begin{eqnarray}\label{Def:PNF}
		\varphi ::= \top \mid \mu \mid \neg \mu \mid \varphi_1 \wedge \varphi_2 \mid \varphi_1 \vee \varphi_2 \mid \phi \mathsf{U}_{\text{I}} \varphi \mid  \mathsf{F}_{\text{I}} \varphi \mid \mathsf{G}_{\text{I}} \phi,
	\end{eqnarray}
	where $\phi::= \top \mid \mu \mid \neg \mu \mid \phi_1 \wedge \phi_2 \mid \phi_1 \vee \phi_2$. Here, $\phi_1, \phi_2$ are formulas of class $\phi$ and $\varphi_1, \varphi_2$ are formulas of class $\varphi$ given in (\ref{Def:PNF}).

\begin{remark}
The STL fragment defined in (\ref{Def:PNF}) includes nested STL formulas of the form $\mathsf{F}_{[a_1, b_1]}\mathsf{G}_{[a_2, b_2]}\phi,  \phi_1\mathsf{U}_{[a_1, b_1]}\mathsf{G}_{[a_2, b_2]}\phi_2$ while excludes nested STL formulas of the form $\mathsf{G}_{[a_1, b_1]}\mathsf{F}_{[a_2, b_2]}\phi, (\mathsf{G}_{[a_1, b_1]}\phi_1)\mathsf{U}_{[a_2, b_2]}\phi_2$. The reason is that  according to the semantics of STL, nested STL formulas like $\mathsf{G}_{[a_1, b_1]}\mathsf{F}_{[a_2, b_2]}\phi$ and $(\mathsf{G}_{[a_1, b_1]}\phi_1)\mathsf{U}_{[a_2, b_2]}\phi_2$ require parallel monitoring of their arguments $\mathsf{F}_{[a_2, b_2]}\phi$ and $\mathsf{G}_{[a_1, b_1]}\phi_1$ within the encoded time intervals of the temporal operators $\mathsf{G}_{[a_1, b_1]}$ and $\mathsf{U}_{[a_2, b_2]}$, respectively.
		Nevertheless, we note that the fragment (\ref{Def:PNF}) is more general than most of the fragments considered in the literature studying online control synthesis, e.g., \citet{lindemann2018,buyukkocak2022control}. Such fragment (\ref{Def:PNF}) is expressive enough to specify a large number of robotic tasks, e.g., time-constrained reachability, supply-delivery, and safety. 
\end{remark}

The problem under consideration is formulated as follows.

\begin{problem}[Online control synthesis]\label{problem2}
	Consider the system (\ref{x0}) and an STL task $\varphi$ in (\ref{Def:PNF}). For an initial state $x_0$, find, if there exists, a sequence of  control inputs $\bm{\nu}=u_0(x_0)u_1(x_1)\ldots u_k(x_k)\ldots$ such that the resulting trajectory $\bm{x}=x_0x_1\ldots x_k\ldots$ satisfies $\varphi$.
\end{problem}

\begin{remark}
Note that the objective of Problem~\ref{problem2} is not to synthesize a closed-form control policy $\bm{\nu}$, which is in general computationally intractable for systems with  continuous spaces.  Instead, we aim at finding online a sequence of feedback control inputs in a way that is similar to  receding horizon control.
\end{remark}

The key idea to solve Problem~\ref{problem2} is as follows. We first transform the STL formula to an alternative tree-based representation, which we call tube-based temporal logic tree (tTLT), by leveraging  reachability analysis, as detailed in Section~\ref{Sec:tTLT}.   There exists a semantic connection between the STL formula and the corresponding tTLT, thanks to the reachability analysis, which is explained in Section~\ref{Sec:STLtTLT}.  Based on this fact, we can
	perform control synthesis over the tTLT, instead of the STL formula. An online control synthesis algorithm is provided in Section~\ref{Sec:Controlsetsynthesis}.

\section{Real-time STL semantics and tube-based temporal logic tree}\label{Sec:tTLT}

In this section, a real-time version of STL semantics and a notion of tTLT are proposed. The real-time STL semantics establishes the satisfaction relation between a real-time signal and the STL formula. Based on this real-time semantics, we then propose the tTLT using the close connection between STL and reachability analysis.

\subsection{Real-time STL semantics}

Before proceeding, the following definition is required.

\begin{definition}[Suffix and Completions]
	Given a discrete-time signal $\bm{x}=x_0x_1\ldots$, we say that a partial signal $\bm{s}=s_{l}s_{l+1}\ldots, l\in \mathbb{N}$, is a \emph{suffix} of the signal $\bm{x}$ if $\forall k'\ge l, s_{k'}=x_{k'}$. The set of \emph{completions} of a partial signal $\bm{s}$, denoted by $C(\bm{s})$, is given by
	\begin{equation*}
		C(\bm{s}):=\{\bm{x}: \bm{s} \ \text{is a suffix of}\ \bm{x}\}.
	\end{equation*}
\end{definition}

Given a time instant $t_k$ and a time interval $[a, b]$, define $t_k+[a, b]:=[t_k+a, t_k+b].$
The real-time STL semantics is defined as follows.

\begin{definition}\label{realtimeSTLsemantic}
Let $t_k$ be the starting time of any STL formula $\varphi$ to be evaluated. Given a partial signal $\bm{s}=s_{l}s_{l+1}\ldots$ starting from time instant $t_{l}\ge t_k$, the real-time satisfaction of $\varphi$ with respect to the partial signal $\bm{s}$, denoted by $(\bm{s}, t_k, t_l) \mid\asymp \varphi$, is recursively defined by Eq.~(\ref{Eq:realSTLsemantic}).
	\begin{figure*}[t]
		\begin{subequations}\label{Eq:realSTLsemantic}
			\begin{eqnarray}
				(\bm{s}, t_k, t_{l}) \mid\asymp \mu &\Leftrightarrow &
				g_\mu(\bm{s}(t_{k}))\ge 0, \quad t_{l}\in t_k+[0,||\mu||];\\
				(\bm{s}, t_k, t_{l}) \mid\asymp  \neg  \varphi &\Leftrightarrow &\neg((\bm{s}, t_{k}, t_l) \mid\asymp \varphi), \quad t_{l}\in t_k+[0,||\varphi||];\\
				(\bm{s}, t_k, t_{l}) \mid\asymp \varphi_1 \wedge \varphi_2 &\Leftrightarrow& (\bm{s}, t_k, t_{l}) \mid\asymp \varphi_1 \wedge  (\bm{s}, t_k, t_{l}) \mid\asymp \varphi_2, \quad t_{l}\in t_k+[0,||\varphi_1 \wedge \varphi_2||];\\
				(\bm{s}, t_k, t_{l}) \mid\asymp \varphi_1 \mathsf{U}_{[a, b]} \varphi_2 &\Leftrightarrow & \nonumber \\
				&&\hspace{-3.5cm}\begin{cases}
					\exists t_{k'}\in [\max\{t_k+a, t_l\}, t_k+b] \ \text{s.t.}\ (\bm{s}, t_{k'}, t_l) \mid\asymp \varphi_2, & \mbox{if } \|\varphi_2\|=0, \\
					\exists t_{k'}\in [t_k+a, t_k+b] \ \text{s.t.}\ (\bm{s}, t_{k'}, t_l) \mid\asymp \varphi_2, & \mbox{otherwise},
				\end{cases} \nonumber\\
				&&\hspace{-3.5cm}\wedge \ \text{if} \ t_l\le t_{k'}, \forall t_{k''}\in [t_{l}, t_{k'}], (\bm{s}, t_{k''}, t_l) \mid\asymp \varphi_1, \quad t_{l}\in t_k+[0, \|\varphi_1 \mathsf{U}_{[a, b]} \varphi_2\|].
			\end{eqnarray}
		\end{subequations}
		\hrule
	\end{figure*}
\end{definition}
The real-time satisfaction relation $(\bm{s}, t_k, t_l) \mid\asymp \varphi$ suggests that the partial signal $\bm{s}$ is the suffix of a satisfying trajectory that starts from $t_k$, \textit{i.e.,}
\begin{equation*}
	(\bm{s}, t_k, t_l) \mid\asymp \varphi\Leftarrow \exists \bm{x}\in C(\bm{s}), (\bm{x}, t_k)\vDash \varphi.
\end{equation*}

Using the induction rule, one can define the real-time STL semantics for ``disjunction" $\varphi_1\vee \varphi_2$, ``eventually" $\mathsf{F}_{[a, b]} \varphi$, and ``always" $\mathsf{G}_{[a, b]} \varphi$.

In parallel with Definitions~\ref{Def:satisfibility}  and  \ref{Def:feasibility}, we define the STL satisfibility given a  partial signal as follows.

\begin{definition}\label{real-timefeasibility}
	Consider the deterministic system (\ref{x1}) and the STL formula $\varphi$. We say $\varphi$ is satisfiable from the state $x_k$ at time $t_k$ if there exists a control policy $\bm{\nu}\in \mathcal{U}_{\ge k}$ such that
	\begin{equation*}\label{satisfy}
		({\bm{x}}_{x_k}^{\bm{\nu}}, t_0, t_k) \mid\asymp \varphi.
	\end{equation*}
\end{definition}

\begin{definition}\label{real-timerobustfeasibility}
	Consider the uncertain system (\ref{x0}) and the STL formula $\varphi$. We say $\varphi$ is robustly satisfiable from the state $x_k$ at time $t_k$ if there exists a control policy $\bm{\nu}\in \mathcal{U}_{\ge k}$ such that
	\begin{equation*}
		({\bm{x}}_{x_k}^{\bm{\nu}, \bm{w}}, t_0, t_k) \mid\asymp \varphi, \forall \bm{w}\in \mathcal{W}_{\ge k}.
	\end{equation*}
\end{definition}

Note that when $t_k=t_0$, Definitions \ref{real-timefeasibility} and \ref{real-timerobustfeasibility} degenerate to Definitions \ref{Def:satisfibility} and \ref{Def:feasibility}, respectively. Given an STL formula $\varphi$, let
\begin{equation}\label{robustsatisfyset}
	\mathbb{S}_{\varphi}(t_k):=\{x_k\in \mathbb{R}^n| \text{$\varphi$ is (robustly) satisfiable from $x_k$ at $t_k$}\}
\end{equation}
denote the set of states from which $\varphi$ is robustly satisfiable at $t_k$. Then, we have the following results.

\begin{proposition}\label{prop1}
	Consider the system (\ref{x0}) and predicates $\mu, \mu_1, \mu_2$. Then, one has
	\begin{itemize}
		\item $\mathbb{S}_{\mu_1 \mathsf{U}_{[a, b]} \mu_2}(t_k)= \mathcal{R}^M(\mathbb{R}^n, \mathbb{S}_{\mu_2},\mathbb{S}_{\mu_1}, [a, b], k)$;
		\item $\mathbb{S}_{\mathsf{F}_{[a, b]} \mu_1}(t_k)= \mathcal{R}^M(\mathbb{R}^n, \mathbb{S}_{\mu_1},\mathbb{R}^n, [a, b], k)$;
		\item $\mathbb{S}_{\mathsf{G}_{[a, b]} \mu_1}(t_k)=  \overline{\mathcal{R}^m(\mathbb{R}^n, \overline{\mathbb{S}_{\mu_1}}, [a, b], k)}$,
	\end{itemize}
	where $\mathbb{S}_{\varphi_1}$ and $\mathbb{S}_{\varphi_2}$ are defined in (\ref{initialsatisfiableset}).
\end{proposition}

\begin{proof}
	The proof follows from Definitions \ref{Def:maxreachset},  \ref{Def:minreachset}, and \ref{realtimeSTLsemantic}.
	\qed
\end{proof}

\begin{proposition}\label{prop2}
	Consider the system (\ref{x0}) and STL formulas $\varphi_1, \varphi_2$. If $\varphi_1$ and $\varphi_2$ contain no logical operators $\wedge$ and $\vee$, then one has
	\begin{itemize}
		\item $\mathbb{S}_{\varphi_1\wedge \varphi_2}(t_k)\subseteq \mathbb{S}_{\varphi_1}(t_k) \cap \mathbb{S}_{\varphi_2}(t_k)$;
		\item $\mathbb{S}_{\varphi_1\vee \varphi_2}(t_k)\supseteq \mathbb{S}_{\varphi_1}(t_k) \cup \mathbb{S}_{\varphi_2}(t_k)$;
	\end{itemize}
	where $\mathbb{S}_{\varphi_1}(t_k)$ and $\mathbb{S}_{\varphi_2}(t_k)$ are defined in (\ref{robustsatisfyset}).
\end{proposition}

\begin{proof}
	Assume that $x_k\in \mathbb{S}_{\varphi_1 \wedge \varphi_2}(t_k)$. According to Definition \ref{realtimeSTLsemantic} and (\ref{robustsatisfyset}), one has that there exists a control policy $\bm{\nu}\in \mathcal{U}_{\ge k}$ such that
	\begin{eqnarray*}
		&&(\bm{x}_{x_k}^{\bm{\nu}, \bm{w}}, t_0, t_k)  \mid\asymp \varphi_1,  \forall \bm{w}\in \mathcal{W}_{\ge k} \\
		&& \wedge \; (\bm{x}_{x_k}^{\bm{\nu}, \bm{w}}, t_0, t_k) \mid\asymp \varphi_2, \forall \bm{w}\in \mathcal{W}_{\ge k}.
	\end{eqnarray*}
	That is, $x_k\in \mathbb{S}_{\varphi_1}(t_k), x_k\in \mathbb{S}_{\varphi_2}(t_k)$. Thus, $x_k\in \mathbb{S}_{\varphi_1\wedge \varphi_2}(t_k) \Rightarrow x_k\in \mathbb{S}_{\varphi_1}(t_k) \cap \mathbb{S}_{\varphi_2}(t_k)$. The other direction may not hold because it could happen that for a state $x_k$, there exist two control policies $\bm{\nu}_1, \bm{\nu}_2 \in \mathcal{U}_{\ge k}$ such that $(\bm{x}_{x_k}^{\bm{\nu}_1, \bm{w}}, t_0, t_k) \mid\asymp \varphi_1, (\bm{x}_{x_k}^{\bm{\nu}_2, \bm{w}}, t_0, t_k) \mid\asymp \varphi_2, \forall \bm{w}\in \mathcal{W}_{\ge k}$ (\textit{i.e.,} $x_k\in \mathbb{S}_{\varphi_1}(t_k)\cap \mathbb{S}_{\varphi_2}(t_k)$). However, there is no control policy which ensures the robust satisfaction of $\varphi_1 \wedge \varphi_2$ at $t_k$.
	
	Assume now that $x_k\in \mathbb{S}_{\varphi_1}(t_k)$, then one has that there exists a control policy $\bm{\nu}\in \mathcal{U}_{\ge k}$ such that $(\bm{x}_{x_k}^{\bm{\nu}, \bm{w}}, t_0, t_k) \mid\asymp \varphi_1, \forall \bm{w}\in \mathcal{W}_{\ge k}$. Moreover, according to STL syntax, one further has
	$(\bm{x}_{x_k}^{\bm{\nu}, \bm{w}}, t_0, t_k) \mid\asymp \varphi_1\vee \varphi_2, \forall \bm{w}\in \mathcal{W}_{\ge k}$.
	That is, $x_k\in \mathbb{S}_{\varphi_1}(t_k)\Rightarrow x_k\in \mathbb{S}_{\varphi_1\vee\varphi_2}(t_k)$. Similarly, one can also get $x_k\in \mathbb{S}_{\varphi_2}(t_k)\Rightarrow x_k\in \mathbb{S}_{\varphi_1\vee\varphi_2}(t_k)$. Therefore, $x_k\in \mathbb{S}_{\varphi_1}(t_k)\cup \mathbb{S}_{\varphi_2}(t_k)\Rightarrow x_k\in \mathbb{S}_{\varphi_1\vee\varphi_2}(t_k)$. The other direction may not hold because it could happen that there exists no state such that either $\varphi_1$ or $\varphi_2$ is robustly satisfiable from at $t_k$, \textit{i.e.,} $\mathbb{S}_{\varphi_1}(t_k)= \emptyset$,  $\mathbb{S}_{\varphi_2}(t_k)=\emptyset$ and thus $\mathbb{S}_{\varphi_1}(t_k)\cup \mathbb{S}_{\varphi_2}(t_k)=\emptyset$. However, there exists a state $x_k^*$ from which there exists a control policy $\bm{\nu}\in \mathcal{U}_{\ge k}$ such that
	\begin{eqnarray*}
		&&\hspace{-0.2cm}( \bm{x}_{x_k^*}^{\bm{\nu}, \bm{w}_1}, t_0, t_k) \mid\asymp \varphi_1, \forall \bm{w}_1\in \mathcal{W}_1 \\
		&& \wedge \; ( \bm{x}_{x_k^*}^{\bm{\nu}, \bm{w}_2}, t_0, t_k) \mid\asymp \varphi_2, \forall \bm{w}_2\in \mathcal{W}_{\ge k}\setminus \mathcal{W}_1,
	\end{eqnarray*}
	where $\mathcal{W}_1\subset \mathcal{W}_{\ge k}$. In this case, one has $x_k^*\in \mathbb{S}_{\varphi_1\vee \varphi_2}(t_k)$.
 \qed
\end{proof}

It is implied from Propositions \ref{prop1} and \ref{prop2} that the real-time satisfiable set of the STL formula can be inferred by set operations and reachability analysis, which makes it reasonable to develop the tTLT, a tree structure consisting of reachable tubes and operators. In the following section, we will detail the definition of tTLT and how to construct an tTLT from a given STL formula using reachability analysis.

\subsection{Tube-based temporal logic tree and its construction}
An tTLT is a variant of the TLT proposed in the recent work \citep{gao2020} for LTL formulas.  Due to the time-dependent essence of STL formulas, the reachable sets in the TLT are replaced with the reachable tubes in the tTLT, which can explicitly incorporate the time constraints in the STL formulas. The intuition of the tTLT is that it indicates how a state trajectory should evolve in order to satisfy the time constrains embedded in an STL formula. In the following, a formal definition of the tTLT is introduced.

\begin{definition}
	An tTLT is a tree for which the next holds:
	\begin{itemize}
		\item each node is either a \emph{tube} node that maps from the nonnegative time axis, \textit{i.e.,} $\mathbb{R}_{\ge 0}$, to the subset of $\mathbb{R}^{n}$, or an \emph{operator} node that belongs to $\{\wedge, \vee, \mathsf{U}_{\text{I}}, \mathsf{F}_{\text{I}}, \mathsf{G}_{\text{I}}\}$;
		\item the root node and the leaf nodes are \emph{tube} nodes;
		\item if a \emph{tube} node is not a leaf node, its unique child is an \emph{operator} node;
		\item the children of any \emph{operator} node are \emph{tube} nodes.
	\end{itemize}
\end{definition}

The following result shows how to construct an tTLT for any given STL formula using reachability analysis.

\begin{theorem}\label{thm1}
	For the system (\ref{x0}) and every STL formula $\varphi$ in (\ref{STLdef}), an tTLT, denoted by $\mathcal{T}_\varphi$, can be constructed from $\varphi$ through the reachability operators $\mathcal{R}^M$ and $\mathcal{R}^m$.
\end{theorem}

\begin{proof}
	We follow three steps to construct an tTLT.
	
	\emph{Step 1:} Rewrite the STL formula $\varphi$ into the equivalent positive normal form (PNF). It has been proven in \citet{sadraddini2015} that each STL formula has an equivalent
	STL formula in PNF (\textit{i.e.,} negations only occur adjacent to predicates), which can be inductively defined as
	\begin{eqnarray*}
		\varphi ::= \top \mid \mu \mid \neg \mu \mid \varphi_1 \wedge \varphi_2 \mid \varphi_1 \vee \varphi_2 \mid \varphi_1 \mathsf{U}_{\text{I}} \varphi_2 \mid  \mathsf{F}_{\text{I}} \varphi_1 \mid \mathsf{G}_{\text{I}} \varphi_1.
	\end{eqnarray*}
	
	\emph{Step 2:} For each predicate $\mu$ or its negation $\neg \mu$, construct the tTLT with only one tube node $\mathbb{X}_{\mu}=\{x: g_{\mu}(x)\ge  0\}$ or $\overline{\mathbb{S}_{\mu}}$. The tTLT of $\top$ or $\bot$ has only one tube node, which is $\mathbb{R}^n$ or $\emptyset$.
	
	\emph{Step 3:} Following the induction rule to construct the tTLT $\mathcal{T}_{\varphi}$. More specifically, we will show that given STL formulas $\varphi_1$ and $\varphi_2$, if the tTLTs can be constructed from $\varphi_1$ and $\varphi_2$, then the tTLTs can be constructed from $\varphi_1 \wedge \varphi_2$, $\varphi_1 \vee \varphi_2$, $\varphi_1\mathsf{U}_{[a, b]} \varphi_2$, $\mathsf{F}_{[a, b]} \varphi_1$, and $\mathsf{G}_{[a, b]} \varphi_1$.
	
	\emph{Case 1}: Boolean operators $\wedge$ and $\vee$. Consider two STL formulas $\varphi_1, \varphi_2$ and their corresponding tTLTs $\mathcal{T}_{\varphi_1}, \mathcal{T}_{\varphi_2}$. The root nodes of $\mathcal{T}_{\varphi_1}$ and $\mathcal{T}_{\varphi_2}$ are denoted by $\mathbb{X}_{\varphi_1}(t_k)$ and $\mathbb{X}_{\varphi_2}(t_k)$, respectively. The tTLT $\mathcal{T}_{\varphi_1 \wedge \varphi_2}$ ($\mathcal{T}_{\varphi_1 \vee \varphi_2}$) can be constructed by connecting $\mathbb{X}_{\varphi_1}(t_k)$ and $\mathbb{X}_{\varphi_2}(t_k)$ through the operator node $\wedge$ ($\vee$) and taking the intersection (or union) of the two root nodes, \textit{i.e.,} $\mathbb{X}_{\varphi_1}(t_k) \cap \mathbb{X}_{\varphi_2}(t_k)$ ($\mathbb{X}_{\varphi_1}(t_k) \cup \mathbb{X}_{\varphi_2}(t_k)$), to be the root node. An illustrative diagram for $\varphi_1 \wedge \varphi_2$ is given in Figure~\ref{wedge}.
	
	\begin{figure}[H]
		\centering
		\subfigure{
			\includegraphics[width=0.4\textwidth]{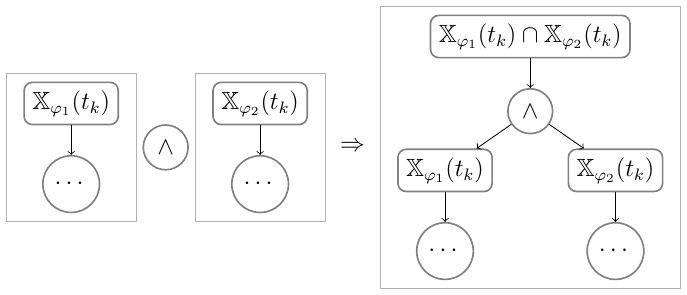}}
		\caption{\footnotesize Illustrative diagram of construction tTLT for $\varphi_1 \wedge \varphi_2$.}
		\label{wedge}
	\end{figure}
	
	\emph{Case 2}: Until operator $\mathsf{U}_{[a, b]}$. Consider two STL formulas $\varphi_1, \varphi_2$ and their corresponding tTLTs $\mathcal{T}_{\varphi_1}, \mathcal{T}_{\varphi_2}$. The root nodes of $\mathcal{T}_{\varphi_1}$ and $\mathcal{T}_{\varphi_2}$ are denoted by $\mathbb{X}_{\varphi_1}(t_k)$ and $\mathbb{X}_{\varphi_2}(t_k)$, respectively. In addition, the leaf nodes of $\mathcal{T}_{\varphi_1}$ are denoted by $\mathbb{Y}_{\varphi_1}^1(t_k), \cdots, \mathbb{Y}_{\varphi_1}^{N}(t_k)$, where $N$ is the total number of leaf nodes of $\mathcal{T}_{\varphi_1}$.
	The tTLT $\mathcal{T}_{\varphi_1 \mathsf{U}_{[a, b]} \varphi_2}$ can be constructed by the following steps: 1) replace each leaf node $\mathbb{Y}_{\varphi_1}^i(t_k)$ by $\mathcal{R}^M(\mathbb{R}^n, \mathbb{X}_{\varphi_2}(t_0), \mathbb{Y}_{\varphi_1}^i(t_0), [a, b],k)$; 2) update $\mathcal{T}_{\varphi_1}$ from the leaf nodes to the root node with the new leaf nodes; and 3) connect each leaf node of the updated $\mathcal{T}_{\varphi_1}$ and the root node of $\mathcal{T}_{\varphi_2}$, \textit{i.e.,} $\mathbb{X}_{\varphi_2}(t_k)$, with the operator node $\mathsf{U}_{[a, b]}$. One illustrative diagram for $\mathsf{U}_{[a, b]}$ is given in Figure~\ref{until}.
	\begin{figure*}
		\centering
		\subfigure{
			\includegraphics[width=0.8\textwidth]{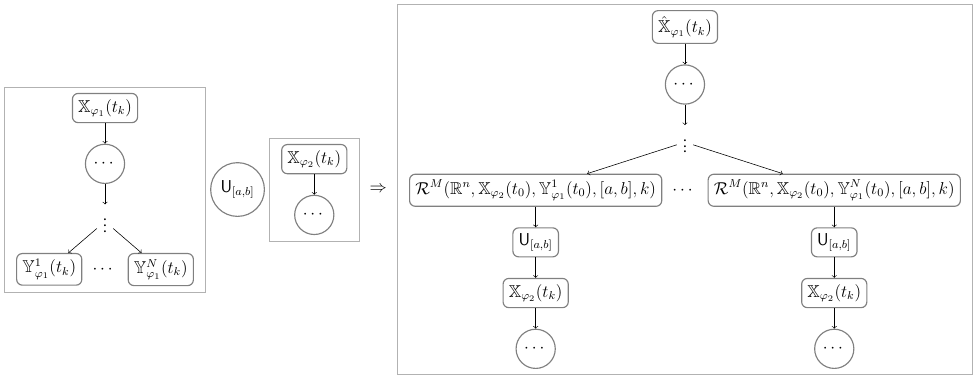}}
		\caption{\footnotesize Illustrative diagram of construction tTLT for $\varphi_1\mathsf{U}_{[a, b]} \varphi_2$.}
		\label{until}
	\end{figure*}
	
	\emph{Case 3}: Eventually and always operators $\mathsf{F}_{[a, b]}$ and $\mathsf{G}_{[a, b]}$. Consider an STL formula $\varphi_1$ and its corresponding tTLT $\mathcal{T}_{\varphi_1}$. The root node of $\mathcal{T}_{\varphi_1}$ is given by $\mathbb{X}_{\varphi_1}(t_k)$. The tTLT $\mathcal{T}_{\mathsf{F}_{[a, b]}{\varphi_1}}$ ($\mathcal{T}_{\mathsf{G}_{[a, b]}{\varphi_1}}$) can be constructed by connecting  $\mathbb{X}_{\varphi_1}(t_k)$ through the operator $\mathsf{F}_{[a, b]}$ ($\mathsf{G}_{[a, b]}$) and making the tube $\mathcal{R}^M(\mathbb{R}^n, \mathbb{X}_{\varphi_1}(t_0), \mathbb{R}^n, [a, b],k)$ ($\overline{\mathcal{R}^m(\mathbb{R}^n, \overline{\mathbb{X}_{\varphi_1}(t_0)}, [a, b],k)}$) the root node. An illustrative diagram for $\mathsf{G}_{[a, b]}$ is given in Figure~\ref{always}.
 \qed

	\begin{figure}
		\centering
		\subfigure{
			\includegraphics[width=0.4\textwidth]{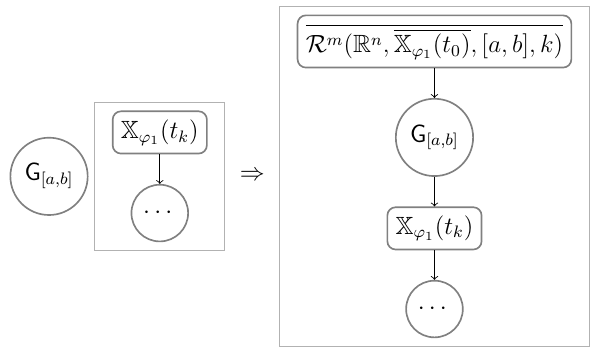}}
		\caption{\footnotesize Illustrative diagram of construction tTLT for $\mathsf{G}_{[a, b]} \varphi_1$.}
		\label{always}
	\end{figure}
\end{proof}

Based on Theorem \ref{thm1}, Algorithm 1 is designed for the construction of tTLT $\mathcal{T}_\varphi$. It takes the syntax tree of the STL formula $\varphi$ as input. For an STL formula, the nodes of its syntax tree are either predicate or operator nodes. More specifically, all the leaf nodes are predicates and all other nodes are operators.

\begin{algorithm}\label{alg1}
	\caption{\textit{tTLTConstruction}}
	\begin{algorithmic}[1]
		\Require the syntax tree of STL formula $\varphi$.
		\Ensure the tTLT $\mathcal{T}_{\varphi}$.
		\For {each leaf node $\mu$ (or $\neg \mu$) of the syntax tree},
		\State {Replace $\mu$ (or $\neg \mu$) by $\mathbb{S}_{\mu}$ (or $\mathbb{S}_{\neg\mu}$)},
		\EndFor
		\For {each operator node of the syntax tree through a bottom-up traversal,}
		\State Construct $\mathcal{T}_{\varphi}$ according to Theorem \ref{thm1},
		\EndFor
	\end{algorithmic}
\end{algorithm}

Let us use the following example to show how to construct the tTLT.

\begin{figure}
	\centering
	\subfigure{
		\includegraphics[width=0.4\textwidth]{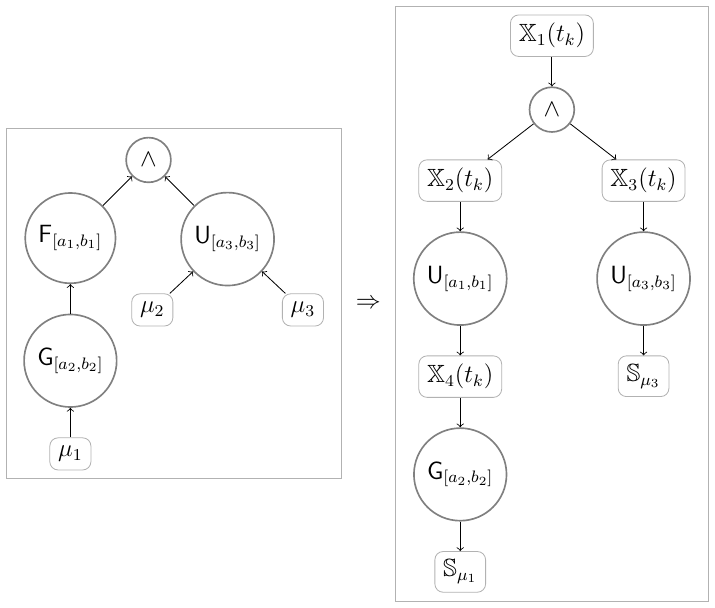}}
	\caption{\footnotesize Example \ref{example1}: syntax tree (left) and tTLT (right) for $\varphi=\mathsf{F}_{[a_1, b_1]}\mathsf{G}_{[a_2, b_2]}\mu_1 \wedge \mu_2 \mathsf{U}_{[a_3, b_3]}\mu_3$. Recall that $\mathsf{F}_{[a, b]} \varphi=\top \mathsf{U}_{[a, b]} \varphi$.}\label{Fig:tTLT}
\end{figure}

\begin{example}\label{example1}
	Consider the formula $\varphi=\mathsf{F}_{[a_1, b_1]}\mathsf{G}_{[a_2, b_2]}\mu_1 \wedge \mu_2 \mathsf{U}_{[a_3, b_3]}\mu_3$, where $\mu_i, i=\{1,2,3\}$ are predicates. The syntax tree of $\varphi$ is shown on the left-hand side of Figure \ref{Fig:tTLT}. The corresponding tTLT for $\varphi$ (constructed using Algorithm 1) is shown on the right-hand side of Figure \ref{Fig:tTLT}, where
	\begin{eqnarray*}
		&& \mathbb{X}_4(t_k)=\overline{\mathcal{R}^m(\mathbb{R}^n, \overline{\mathbb{S}_{\mu_1}}, [a_2, b_2], k)},\\
		&& \mathbb{X}_3(t_k)=\mathcal{R}^M(\mathbb{R}^n, \mathbb{S}_{\mu_3}, \mathbb{S}_{\mu_2}, [a_3, b_3], k),\\
		&& \mathbb{X}_2(t_k)=\mathcal{R}^M(\mathbb{R}^n, \mathbb{X}_{4}(t_0),\mathbb{R}^n, [a_1, b_1], k),\\
		&&\mathbb{X}_1(t_k)=\mathbb{X}_2(t_k)\cap \mathbb{X}_3(t_k).
	\end{eqnarray*}
\end{example}

\begin{remark}\label{TLT:compucomlex}
	Given an STL formula $\varphi$ in positive normal form, let $N$ denote the number of Boolean operators and $M$ the number of temporal operators contained in $\varphi$. Let $\mathcal{T_\varphi}$ be the tTLT corresponds to $\varphi$. Then, $\mathcal{T_\varphi}$ has at most $2N$ number of complete paths. In addition, each complete path has at most $2(N+M)+1$ number of nodes, out of which at most $N+M$ are non-root tube nodes. Thus, one can conclude that $\mathcal{T_\varphi}$ contains at most $4N(N+M)+1$ number of nodes, out of which at most $2N(N+M)+1$ number of tube nodes.
\end{remark}

\section{Semantic Connection between STL and tTLT}\label{Sec:STLtTLT}

In this section, the semantic connection between an STL formula and its corresponding tTLT is derived.
Before that, we first define the \emph{complete path} and its \emph{segment}.

\begin{definition}
	A \emph{complete path} $\bm{p}$ of an tTLT is a path that starts from the root node and ends at a leaf node. It can be encoded in the form of $\bm{p}=\mathbb{X}_0\Theta_1\mathbb{X}_1\Theta_2\ldots \Theta_{N_f} \mathbb{X}_{N_f}$, where $N_f$ is the number of operator nodes contained in the complete path,  $\mathbb{X}_i:  \mathbb{R}_{\ge 0}\to 2^{\mathbb{R}^n}, \forall i\in \{0,1,\ldots,N_f\}$ represent tube nodes, and $\Theta_{j}\in \{\wedge, \vee, \mathsf{U}_{\text{I}}, \mathsf{F}_{\text{I}},   \mathsf{G}_{\text{I}}\}, \forall j\in \{1,\ldots,N_f\}$ represent operator nodes. Any subsequence of a complete path is called a \emph{segment} of the complete path.
\end{definition}

Now, we define the \emph{maximal temporal segment} for an tTLT, which plays an important role when simplifying the tTLT.

\begin{definition}\label{Def:minBoolfrag}
	A \emph{maximal temporal segment (MTS)} of a complete path of the tTLT is one of the following types of segment:
	\begin{itemize}
		\item[1)] a segment from the root node to the parent of the first Boolean operator node  ($\wedge$ or $\vee$);
		\item[2)] a segment from one child of one Boolean operator node to the parent of the next Boolean operator node;
		\item[3)] a segment from one child of the last Boolean operator node to the leaf node.
	\end{itemize}
\end{definition}

One can conclude from Definition \ref{Def:minBoolfrag} that any MTS starts and ends with a tube node and contains no Boolean operator nodes.

\begin{definition}\label{Def:timecoding}
	A \emph{time coding} of (a complete path of) the tTLT is an assignment of each tube node $\mathbb{X}_i$ of (the complete path of) the tTLT an activation time instant $t_{\kappa_i}, \kappa_i\in \mathbb{N}$.
\end{definition}

Now, we further define the satisfaction relation between a trajectory $\bm{x}$ and a complete path of the tTLT.

\begin{definition}\label{Def:PathSaf}
	Consider a trajectory $\bm{x}:=x_0x_1\ldots$ and a complete path $\bm{p}=\mathbb{X}_0\Theta_1\mathbb{X}_1\Theta_2\ldots \Theta_{N_f} \mathbb{X}_{N_f}$. We say \emph{$\bm{x}$ satisfies $\bm{p}$}, denoted by $\bm{x} \vDash \bm{p}$, if there exists a time coding for $\bm{p}$ such that
	\begin{itemize}
		\item[i)] if $\Theta_i\in \{\wedge, \vee\}$, then $t_{\kappa_i}=t_{\kappa_{i-1}}$;
		\item[ii)] if $\Theta_i=\mathsf{U}_{\text{I}}$, then $t_{\kappa_i}\in t_{\kappa_{i-1}}+{\text{I}}$;
		\item[iii)] if $\Theta_i=\mathsf{G}_{\text{I}}$, then $t_{\kappa_i}=\argmax_{t_k}\{t_k\in t_{\kappa_{i-1}}+{\text{I}}\}$;
	\end{itemize}
	and
	\begin{itemize}
		\item[iv)] $x_{k}\in \mathbb{X}_{i}(t_{k-{\kappa_{i}}}), \forall k\in [{\kappa_{i}}, {\kappa_{i+1}}], i=0, \ldots, N_f-1$;
		\item[v)] $x_{\kappa_{N_f}}\in \mathbb{X}_{N_f}(t_{0})$.
	\end{itemize}
\end{definition}

\begin{remark}\label{remark4.1}
	From items i)-iii) of Definition \ref{Def:PathSaf}, one has that $t_{\kappa_0}\le t_{\kappa_1}\le \cdots \le t_{\kappa_{N_f}}$. This means that if a
	trajectory $\bm{x}\vDash \bm{p}$, it must visit each tube node $\mathbb{X}_i$ of the complete path $\bm{p}$ sequentially. In addition, we can further conclude from items iv)-v) that the trajectory $\bm{x}$ has to stay in each tube node $\mathbb{X}_i$ for sufficiently long time steps.
\end{remark}

With Definition \ref{Def:PathSaf}, the satisfaction relation between a trajectory $\bm{x}$ and an tTLT $\mathcal{T}_\varphi$ can be defined as follows.

\begin{definition}\label{Def:TreeSaf}
		Consider a trajectory $\bm{x}$ and an tTLT $\mathcal{T}_\varphi$. We say \emph{$\bm{x}$ satisfies $\mathcal{T}_\varphi$}, denoted by $\bm{x} \vDash \mathcal{T}_\varphi$, if  there exists a time coding $\{t_{\kappa_i}\}$ for $\mathcal{T}_\varphi$ such that the output of Algorithm 2 is ${\rm true}$.
	\end{definition}
	
	The central idea of Algorithm 2 is to check the Boolean relation among sub-formulas of a given STL formula $\varphi$. For instance, assume $\varphi=\wedge_{i=1}^n \varphi_i$, where each $\varphi_i, \forall i=1, \cdots, n$ contains no Boolean operators. Then one can get from Algorithm 1 that $\mathcal{T}_\varphi$ has $n$ complete paths $\bm{p}_i, i=1, \cdots, n$, and each $\bm{p}_i$ corresponds to a sub-formula $\varphi_i$. Then Algorithm 2 dictates that $\bm{x} \vDash \mathcal{T}_\varphi$ if and only if $\bm{x}$ satisfies every complete path of $\mathcal{T}_\varphi$. Assume now that $\varphi=\vee_{i=1}^n \varphi_i$, then Algorithm 2 dictates that $\bm{x} \vDash \mathcal{T}_\varphi$ if and only if $\bm{x}$ satisfies at least one complete path of $\mathcal{T}_\varphi$.

\begin{algorithm}
	\caption{\textit{tTLTSatisfaction}}
	\begin{algorithmic}[1]
		\Require a trajectory $\bm{x}$, an tTLT $\mathcal{T}_\varphi$, and a time coding $\{t_{\kappa_i}\}$.
		\Ensure ${\rm true}$ or {\rm{false}}.
		\State $\mathcal{T}_\varphi^c \leftarrow \textit{Compression}(\mathcal{T}_{\varphi})$,
		\For {each complete path $\bm{p}$ of $\mathcal{T}_\varphi$,}
		\If{$\bm{x}\models \bm{p}$}
		\State \parbox[t]{\dimexpr\linewidth-\algorithmicindent}{set the corresponding leaf node of $\bm{p}$ in \\ $\mathcal{T}_\varphi^c$ with ${\rm true}$,}
		\Else
		\State \parbox[t]{\dimexpr\linewidth-\algorithmicindent}{set the corresponding leaf node of $\bm{p}$ in \\ $\mathcal{T}_\varphi^c$ with ${\rm false}$,}
		\EndIf
		\EndFor
		\State set all the non-leaf tube nodes in $\mathcal{T}_\varphi^c$ with ${\rm false}$,
		\State $\textit{Backtracking}(\mathcal{T}_{\varphi}^c)$,
		\State return the root node of $\mathcal{T}_{\varphi}^c$.
	\end{algorithmic}
\end{algorithm}

Algorithm 2 takes as inputs a trajectory $\bm{x}$, an tTLT $\mathcal{T}_\varphi$, and a time coding $\{t_{\kappa_i}\}$, and outputs ${\rm true}$ or ${\rm false}$. It works as follows. Given an tTLT $\mathcal{T}_\varphi$, we first compress it via Algorithm 3 (line 1), in this way the resulting compressed tree $\mathcal{T}_\varphi^c$ contains only Boolean operator nodes and tube nodes. Then for each complete path $\bm{p}$ of $\mathcal{T}_\varphi$, if $\bm{x}\models \bm{p}$, one sets the corresponding leaf node of $\bm{p}$ in $\mathcal{T}_\varphi^c$ (note that $\mathcal{T}_\varphi^c$ and $\mathcal{T}_\varphi$ have the same set of leaf nodes) with ${\rm true}$. Otherwise, one sets the corresponding leaf node of $\bm{p}$ in $\mathcal{T}_\varphi^c$ with ${\rm false}$ (lines 2-8). After that, we set all the non-leaf tube nodes of $\mathcal{T}_\varphi^c$ with ${\rm false}$ (line 9) and the resulting tree becomes a Boolean tree (a tree with Boolean operator and Boolean variable nodes). Finally, we backtrack the Boolean tree $\mathcal{T}_\varphi^c$ using Algorithm 4, and return the root node (lines 10-11).

We further detail the \textit{Compression} algorithm (Algorithm 3) and the \textit{Backtracking} algorithm (Algorithm 4) in the following. Algorithm 3 aims at obtaining a simplified tree with Boolean operator nodes and tube nodes only. To do so, we first encode each MTS in the form of $\mathbb{X}_1\Theta_1\ldots \Theta_{N_f-1} \mathbb{X}_{N_f}$ (line 3), and then replace it with one tube node (line 4). Algorithm 4 takes the compressed tree $\mathcal{T}_\varphi^c$ as an input, and then update the parent of each Boolean operator node through a bottom-up traversal. In Algorithm 4, $\text{PA}(\Theta)$ and $\text{CH}_1(\Theta), \text{CH}_2(\Theta)$ represent the parent node and the two children of the Boolean operator node $\Theta\in \{\wedge, \vee\}$, respectively.

\begin{algorithm}[H]
	\caption{\textit{Compression}}
	\begin{algorithmic}[1]
		\Require an tTLT $\mathcal{T}_\varphi$.
		\Ensure the compressed tree $\mathcal{T}_\varphi^c$.
		\For {each complete path of $\mathcal{T}_\varphi$,}
		\For {each MTS,}
		\State \parbox[t]{\dimexpr\linewidth-\algorithmicindent}{encode the MTS in the form \\ of $\mathbb{X}_1(t_k)\Theta_1\ldots \Theta_{N_f-1} \mathbb{X}_{N_f}(t_k)$,}
		\State replace the MTS with one tube node $\cup_{i=1}^{N_f}\mathbb{X}_{i}$,
		\EndFor
		\EndFor
	\end{algorithmic}
\end{algorithm}

\begin{algorithm}
	\caption{\textit{Backtracking}}
	\begin{algorithmic}[1]
		\Require a compressed tree $\mathcal{T}_\varphi^c$.
		\Ensure the root node of $\mathcal{T}_\varphi^c$.
		\For {each Boolean operator node $\Theta$ of $\mathcal{T}_\varphi^c$ through a bottom-up traversal,}
		\If {$\Theta=\wedge$,}
		\State $\text{PA}(\Theta)\leftarrow\text{PA}(\Theta)\vee (\text{CH}_1(\Theta)\wedge \text{CH}_2(\Theta))$,
		\Else
		\State $\text{PA}(\Theta)\leftarrow\text{PA}(\Theta)\vee (\text{CH}_1(\Theta)\vee \text{CH}_2(\Theta))$,
		\EndIf
		\EndFor
	\end{algorithmic}
\end{algorithm}

\begin{figure}[H]
	\centering
	\subfigure{
		\includegraphics[width=0.4\textwidth]{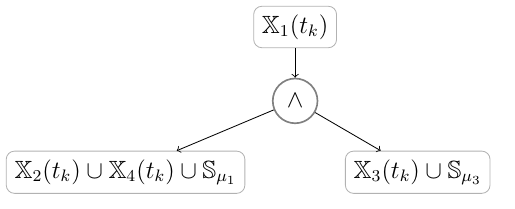}}
	\caption{\footnotesize Example \ref{example1.1}: compressed tree $\mathcal{T}_\varphi^c$, where $\mathcal{T}_\varphi$ is plotted in Figure \ref{Fig:tTLT}.}\label{Fig:comptTLT}
\end{figure}

\begin{example}\label{example1.1}
	Let us continue with Example \ref{example1}. The tTLT $\mathcal{T}_\varphi$ (right of Figure \ref{Fig:tTLT}) contains 2 complete paths,  \textit{i.e.,}
	\begin{equation*}
		\bm{p}_1:=\mathbb{X}_1 \wedge \mathbb{X}_2\mathsf{U}_{[a_1, b_1]}\mathbb{X}_4\mathsf{G}_{[a_2, b_2]}\mathbb{S}_{\mu_1}
	\end{equation*}
	and
	\begin{equation*}
		\bm{p}_2:= \mathbb{X}_1 \wedge\mathbb{X}_3\mathsf{U}_{[a_3, b_3]}\mathbb{S}_{\mu_3}.
	\end{equation*}
	Let
	\begin{equation*}
		\{t_{\kappa_1}, t_{\kappa_2}, t_{\kappa_4}, t_{\kappa_5}\}
	\end{equation*}
	be the time coding of the complete path $\bm{p}_1$, where $t_{\kappa_1}, t_{\kappa_2}, t_{\kappa_4}$, and $t_{\kappa_5}$ are the activation time instants of the tube nodes $\mathbb{X}_1, \mathbb{X}_2, \mathbb{X}_4$, and $\mathbb{X}_5:=\mathbb{S}_{\mu_1}$, respectively. Then, we have according to Definition \ref{Def:PathSaf} that a trajectory $\bm{x}\vDash \bm{p}_1$ if i) $t_{\kappa_1}= t_{\kappa_2}$; ii) $t_{\kappa_4}\in t_{\kappa_2}+[a_1, b_1]$; iii) $t_{\kappa_5}=\argmax_{t_k}\{t_k\in t_{\kappa_{4}}+[a_2, b_2]\}$; iv) $x_0\in \mathbb{X}_1(t_0)$, $x_k\in \mathbb{X}_2(t_{k-\kappa_2}), \forall k\in [\kappa_2, \kappa_4]$, $x_k\in \mathbb{X}_4(t_{k-\kappa_4}), \forall k\in [\kappa_4, \kappa_5]$, and v) $x_{\kappa_5}\in \mathbb{X}_5$.
	
	In addition, the tTLT $\mathcal{T}_\varphi$ contains 3 MTSs, \textit{i.e.,} $\mathbb{X}_1$, $\mathbb{X}_2\mathsf{U}_{[a_1, b_1]}\mathbb{X}_4\mathsf{G}_{[a_2, b_2]}\mathbb{S}_{\mu_1}$, and $\mathbb{X}_3\mathsf{U}_{[a_3, b_3]}\mathbb{S}_{\mu_3}$. The compressed tree $\mathcal{T}_\varphi^c$ is shown in Figure \ref{Fig:comptTLT}. If a trajectory $\bm{x}$ satisfies both of the complete paths $\bm{p}_1$ and $\bm{p}_2$, the output of Algorithm 2 is ${\rm true}$, otherwise, the output is ${\rm false}$.
\end{example}

\begin{definition}(Robust satisfiable tTLT)\label{Def:robustsatistTLT}
	The tTLT $\mathcal{T}_\varphi$ is called robust satisfiable for the system (\ref{x0}) with initial state $x_0$ if there exists a control policy $\bm{\nu}\in \mathcal{U}_{\ge 0}$ such that ${\bm{x}}_{x_0}^{\bm{\nu}, \bm{w}}  \vDash \mathcal{T}_\varphi, \forall \bm{w}\in \mathcal{W}_{\ge 0}$.
\end{definition}

The following theorem provides a formally semantic relation between the STL formula fragment in   (\ref{Def:PNF}) and the corresponding tTLTs. 
	
	\begin{theorem}\label{The:SLTtLTL}
		Consider the uncertain system (\ref{x0}) with initial state $x_0$ and an STL formula $\varphi$ in (\ref{Def:PNF}). Let $\mathcal{T}_\varphi$ be the tTLT corresponding to $\varphi$. Then, one has that $\varphi$ is robustly satisfiable for (\ref{x0}) if  $\mathcal{T}_\varphi$ is robustly satisfiable for (\ref{x0}).
\end{theorem}

\begin{proof}
	From Definitions \ref{Def:feasibility} and \ref{Def:robustsatistTLT}, one has that to prove Theorem \ref{The:SLTtLTL}, it is equivalent to prove
	$\bm{x}_{x_0}^{\bm{\nu}, \bm{w}}\vDash \mathcal{T}_\varphi, \forall \bm{w}\in \mathcal{W}_{\ge 0} \Rightarrow \bm{x}_{x_0}^{\bm{\nu},\bm{w}}\vDash \varphi, \forall \bm{w}\in \mathcal{W}_{\ge 0}$. Given one instance of disturbance signal $\bm{w}$, if one has $\bm{x}_{x_0}^{\bm{\nu}, \bm{w}}\vDash \mathcal{T}_\varphi \Rightarrow \bm{x}_{x_0}^{\bm{\nu},\bm{w}}\vDash \varphi$, then it implies $\bm{x}_{x_0}^{\bm{\nu}}\vDash \mathcal{T}_\varphi, \forall \bm{w}\in \mathcal{W}_{\ge 0}  \Rightarrow \bm{x}_{x_0}^{\bm{\nu},\bm{w}}\vDash \varphi, \forall \bm{w}\in \mathcal{W}_{\ge 0}$. Therefore, it is sufficient to prove
	$\bm{x}_{x_0}^{\bm{\nu}, \bm{w}}\vDash \mathcal{T}_\varphi \Rightarrow \bm{x}_{x_0}^{\bm{\nu},\bm{w}}\vDash \varphi$.
	
	In the following, we will first prove $\bm{x}_{x_0}^{\bm{\nu}, \bm{w}}\vDash \mathcal{T}_\varphi \Leftrightarrow \bm{x}_{x_0}^{\bm{\nu}, \bm{w}}\vDash \varphi$ for
	\begin{itemize}
		\item [\it{i)}] $\top$, predicates $\mu, \neg \mu$, and $\mu_1\wedge \mu_2, \mu_1\vee \mu_2$,
		\item [\it{ii)}] $\mu_1 \mathsf{U}_{[a, b]} \mu_2$, $\mathsf{F}_{[a, b]} \mu_1$, and $\mathsf{G}_{[a, b]} \mu_1$;
		\item [\it{iii)}] $\mu_1 \mathsf{U}_{[a_1, b_1]} \mathsf{G}_{[a_2, b_2]}\mu_2$ and $\mathsf{F}_{[a_1, b_1]} \mathsf{G}_{[a_2, b_2]}\mu_1$;
		\item [\it{iv)}] $\varphi_1 \wedge \varphi_2$;
	\end{itemize}
	where $\varphi_1$ and $\varphi_2$ in item iv) are STL formulas belong to items ii) or iii).
	
	\emph{Case i)}: For $\top$, predicates $\mu, \neg \mu$, and $\mu_1\wedge \mu_2, \mu_1\vee \mu_2$, it is trivial to verify that $\bm{x}_{x_0}^{\bm{\nu}, \bm{w}}\vDash \mathcal{T}_\varphi \Leftrightarrow \bm{x}_{x_0}^{\bm{\nu}, \bm{w}}\vDash \varphi$.
	
	\emph{Case ii)}: We note that the proofs of the three are similar, therefore, in the following, we only consider the case $\varphi=\mu_1 \mathsf{U}_{[a, b]} \mu_2$. The tTLT $\mathcal{T}_{\varphi}$ can be constructed via Algorithm 1, which is shown in Figure \ref{simpleuntil}.
	\begin{figure}[H]
		\centering
		\subfigure{
			\includegraphics[width=0.4\textwidth]{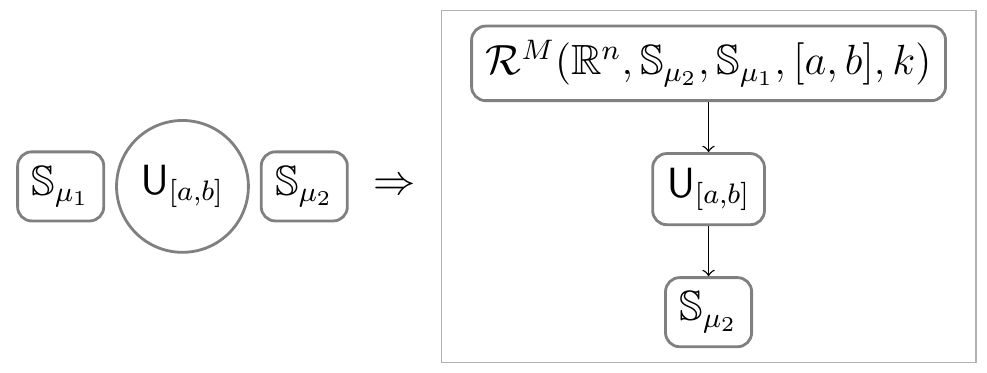}}
		\caption{tTLTs $\mathcal{T}_{\varphi}$ for $\varphi=\mu_1 \mathsf{U}_{[a, b]} \mu_2$.}
		\label{simpleuntil}
	\end{figure}
	Assume that $\bm{x}_{x_0}^{\bm{\nu}, \bm{w}}\vDash \mathcal{T}_\varphi$, then one has from Definition \ref{Def:PathSaf} that $\exists t_{\kappa_1}\in t_0+[a, b], x_{\kappa_1}\in \mathbb{S}_{\mu_2}$ and $\forall k\in [0, {\kappa_1}], x_{k}\in \mathcal{R}^M(\mathbb{R}^n, \mathbb{S}_{\mu_2}, \mathbb{S}_{\mu_1}, [a, b], k)\subseteq \mathbb{S}_{\mu_1}$, which implies $\bm{x}_{x_0}^{\bm{\nu}, \bm{w}}\vDash \varphi$. That is, $\bm{x}_{x_0}^{\bm{\nu}, \bm{w}}\vDash \mathcal{T}_\varphi\Rightarrow \bm{x}_{x_0}^{\bm{\nu}, \bm{w}}\vDash \varphi$. Assume now that $\bm{x}_{x_0}^{\bm{\nu}, \bm{w}}\vDash \varphi$. Then, one has from STL semantics that i) $\exists t_{k'}\in t_0+[a, b], x_{k'}\in \mathbb{S}_{\varphi_2}$ and ii) $\forall t_{k''}\in [t_0, t_{k'}], x_{k''}\in \mathbb{S}_{\varphi_1}$. Moreover, from Definition \ref{Def:maxreachset}, one has that i) and ii) together implies $\forall t_{k''}\in [t_0, t_{k'}], x_{k''}\in \mathcal{R}^M(\mathbb{R}^n, \mathbb{S}_{\mu_2}, \mathbb{S}_{\mu_1}, [a, b], k'')$. Therefore, $\bm{x}_{x_0}^{\bm{\nu}, \bm{w}}\vDash \varphi \Rightarrow\bm{x}_{x_0}^{\bm{\nu}}\vDash \mathcal{T}_\varphi$.
	
	\emph{Case iii)}: We note that the proofs of the two are similar. In the following, we consider the case $\varphi=\mathsf{F}_{[a_1, b_1]} \mathsf{G}_{[a_2, b_2]}\mu_1$. The tTLT $\mathcal{T}_{\varphi}$ can be constructed via Algorithm 1, which is shown in Figure \ref{simpleeventuallyalways}.
	\begin{figure}[H]
		\centering
		\subfigure{
			\includegraphics[width=0.4\textwidth]{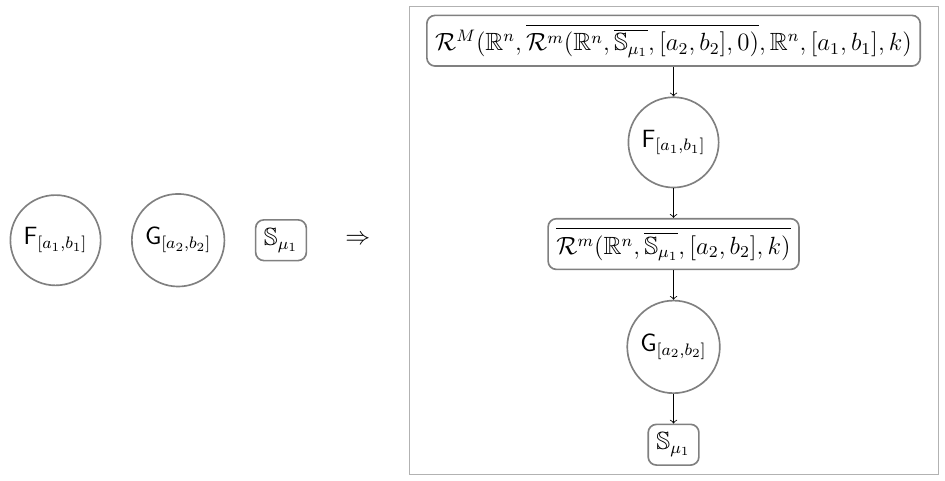}}
		\caption{tTLTs $\mathcal{T}_{\varphi}$ for $\varphi=\mathsf{F}_{[a_1, b_1]} \mathsf{G}_{[a_2, b_2]}\mu_1$.}
		\label{simpleeventuallyalways}
	\end{figure}
	Assume that $\bm{x}_{x_0}^{\bm{\nu}, \bm{w}}\vDash \mathcal{T}_\varphi$, then one has from Definition \ref{Def:PathSaf} that $t_{\kappa_1}\in t_0+[a_1, b_1], t_{\kappa_2}=\argmax_{t_k}\{t_{\kappa_1}+[a_2, b_2]\}$. In addition, $\forall k\in [{\kappa_1}, {\kappa_2}], x_{k}\in \overline{\mathcal{R}^m(\mathbb{R}^n, \overline{\mathbb{S}_{\mu_1}}, [a_2, b_2], k-\kappa_1)}$, which implies $x_k\in \mathbb{S}_{\mu_1}, \forall k\in [{\kappa_1}, {\kappa_2}]$. That is, $\bm{x}_{x_0}^{\bm{\nu}, \bm{w}}\vDash \mathcal{T}_\varphi\Rightarrow \bm{x}_{x_0}^{\bm{\nu}, \bm{w}}\vDash \varphi$. Assume now that $\bm{x}_{x_0}^{\bm{\nu}, \bm{w}}\vDash \varphi$. Then, one has from STL semantics that $\exists t_{k'}\in t_0+[a_1, b_1]$ such that $x_{k''}\in \mathbb{S}_{\mu_1}, \forall t_{k''}\in t_{k'}+[a_2, b_2]$, which implies $\forall t_{k''}\in t_{k'}+[a_2, b_2], x_{k''}\in \overline{\mathcal{R}^m(\mathbb{R}^n, \overline{\mathbb{S}_{\mu_1}}, [a_2, b_2], k''-k')}$. Therefore, $\bm{x}_{x_0}^{\bm{\nu}, \bm{w}}\vDash \varphi \Rightarrow\bm{x}_{x_0}^{\bm{\nu}}\vDash \mathcal{T}_\varphi$.
	
	\emph{Case iv)}: $\varphi=\varphi_1 \wedge \varphi_2$. Assume that $\bm{x}_{x_0}^{\bm{\nu}, \bm{w}}\vDash \mathcal{T}_\varphi$, then one has from Definition \ref{Def:PathSaf} that $\bm{x}_{x_0}^{\bm{\nu}, \bm{w}}\vDash \mathcal{T}_{\varphi_1}$ and $\bm{x}_{x_0}^{\bm{\nu}, \bm{w}}\vDash \mathcal{T}_{\varphi_2}$. Moreover, since $\varphi_1$ and $\varphi_2$ belong to items ii) or iii), then one can conclude from \emph{Case ii)} and \emph{Case iii)} that $\bm{x}_{x_0}^{\bm{\nu}, \bm{w}}\vDash \mathcal{T}_{\varphi_i}\Rightarrow \bm{x}_{x_0}^{\bm{\nu}, \bm{w}}\vDash \varphi_i, i=\{1,2\}$, which implies $\bm{x}_{x_0}^{\bm{\nu}, \bm{w}}\vDash \varphi_1 \wedge \varphi_2$. That is, $\bm{x}_{x_0}^{\bm{\nu}, \bm{w}}\vDash \mathcal{T}_\varphi\Rightarrow \bm{x}_{x_0}^{\bm{\nu}, \bm{w}}\vDash \varphi$. The proof of the other direction is similar and hence omitted.
	
	Then, we prove $\bm{x}_{x_0}^{\bm{\nu}, \bm{w}}\vDash \mathcal{T}_\varphi \Rightarrow \bm{x}_{x_0}^{\bm{\nu}, \bm{w}}\vDash \varphi$ for {\it v)} $\varphi_1 \vee \varphi_2$, where $\varphi_1$ and $\varphi_2$ are STL formulas belong to items ii) or iii).
	
	\emph{Case v)}: $\varphi=\varphi_1 \vee \varphi_2$. The proof of $\bm{x}_{x_0}^{\bm{\nu}, \bm{w}}\vDash \mathcal{T}_\varphi\Rightarrow \bm{x}_{x_0}^{\bm{\nu}, \bm{w}}\vDash \varphi$ is similar to \emph{Case iv)}. The other direction does not hold because for an uncertain system, it is possible that there exists a trajectory $\bm{x}_{x_0}^{\bm{\nu}, \bm{w}}$ such that $\bm{x}_{x_0}^{\bm{\nu}, \bm{w}}\vDash \varphi$, however, the initial state $x_0\notin \mathbb{X}_{\text{root}}^\varphi(t_0)$ (due to Proposition \ref{prop2}), where $\mathbb{X}_{\text{root}}^\varphi$ denotes the root node of $\mathcal{T}_\varphi$. In this case, $\bm{x}_{x_0}^{\bm{\nu}, \bm{w}}$ does not satisfy $\mathcal{T}_\varphi$.
	
	The proof of $\bm{x}_{x_0}^{\bm{\nu}, \bm{w}}\vDash \mathcal{T}_\varphi \Rightarrow \bm{x}_{x_0}^{\bm{\nu}, \bm{w}}\vDash \varphi$ for other STL formulas $\varphi$ in (\ref{Def:PNF}) can be completed inductively by combining \emph{Cases i)-v)}. Therefore, the conclusion follows.
	\qed
\end{proof}

Thanks to the semantic relation between the STL formulas in (\ref{Def:PNF}) and the corresponding tTLT, we  are  able to perform control synthesis over the tTLT, instead of the STL formulas, while preserving the correct-by-construction guarantee.  The details of this control synthesis are provided in the next section. 

\section{Online Control Synthesis}\label{Sec:Controlsetsynthesis}

This section concerns online control synthesis as defined by Problem \ref{problem2}. From Theorems~\ref{The:SLTtLTL} , one can see that to guarantee the satisfaction of the STL formula $\varphi$ in (\ref{Def:PNF}), it is sufficient to find a control policy $\bm{\nu}$ that guarantees the (robust) satisfaction of the corresponding tTLT $\mathcal{T}_{\varphi}$. In the following, control synthesis algorithms are designed such that the tTLT $\mathcal{T}_{\varphi}$ is satisfied based on Definitions \ref{Def:PathSaf} and \ref{Def:TreeSaf}.

\subsection{Definitions and notations}
Before proceeding, the following definitions and notations are needed.

\begin{definition}\label{timehorizon}
	The time horizon $|\Theta|$ of an STL operator $\Theta\in \{\wedge, \vee, \mathsf{U}_{[a, b]}, \mathsf{F}_{[a, b]}, \mathsf{G}_{[a, b]}\}$ is defined as
	\begin{equation*}
		|\Theta|=\begin{cases}
			0, & \mbox{if } \Theta=\{\wedge, \vee\}, \\
			\hat b, & \mbox{if} \; \Theta \in \{\mathsf{U}_{[a, b]}, \mathsf{F}_{[a, b]}, \mathsf{G}_{[a, b]}\},
		\end{cases}
	\end{equation*}
	where $\hat b=\argmax_{t_k}\{a\le t_k\le b\}$.
\end{definition}

\begin{definition}\label{Booleanfragement}
	A segment of the complete path of an tTLT is called \emph{a Boolean segment} if it starts and ends with a tube node and contains only Boolean operator nodes. We say a tube node $\mathbb{X}_j$ is \emph{reachable from $\mathbb{X}_i$ by a Boolean segment} if there exists a Boolean segment that starts with $\mathbb{X}_i$ and ends with $\mathbb{X}_j$.
\end{definition}

\begin{definition}\label{controltree}
	If each node of a tree is either a set node that is a subset of $U$ or an operator node that belongs to $\{\wedge, \vee, \mathsf{U}_{\text{I}}, \mathsf{F}_{\text{I}}, \mathsf{G}_{\text{I}}\}$, then the tree is called a \emph{control tree}.
\end{definition}

Each tube node $\mathbb{X}_i$ of the tTLT $\mathcal{T}_\varphi$ is characterized by the following two parameters:
\begin{itemize}
	\item $t_a(\mathbb{X}_i)$: the activation time of $\mathbb{X}_i$,
	\item $t_h(\mathbb{X}_i)$: the time horizon of $\mathbb{X}_i$, \textit{i.e.,} the time that $\mathbb{X}_i$ is deactivated.
\end{itemize}
Denote by $\mathcal{T}_{\varphi}(t_k)$ the resulting tree of $\mathcal{T}_{\varphi}$ at time instant $t_k$.
It is obtained by fixing the value of each tube node $\mathbb{X}_i$ according to the activation time $t_a(\mathbb{X}_i)$ (\textit{i.e.,} $\mathcal{T}_{\varphi}(t_k)$ contains either set nodes or operator nodes). Let $S_i(t_k)$ be the $i$-th set node of $\mathcal{T}_{\varphi}(t_k)$, where $S_i(t_k)$ corresponds to the tube node $\mathbb{X}_i$. The relationship between $S_i(t_k)$ and $\mathbb{X}_i$ can be described as follows:
\begin{equation}\label{sitk}
	S_i(t_k)=\begin{cases}
		\mathbb{X}_i(t_0), & \mbox{if } t_k\le t_a(\mathbb{X}_i), \\
		\mathbb{X}_i(t_k-t_a(\mathbb{X}_i)), & \mbox{if } t_k> t_a(\mathbb{X}_i).
	\end{cases}
\end{equation}
Moreover, one has that
\begin{equation*}
	t_a(S_i(t_k))=t_a(\mathbb{X}_i), t_h(S_i(t_k))=t_h(\mathbb{X}_i), \forall k\ge 0.
\end{equation*}

At each time instant $t_k$, $\mathcal{T}_{\varphi}(t_k)$ is characterized by
\begin{itemize}
	\item $P(t_k)$: the set which collects all the set nodes of $\mathcal{T}_{\varphi}(t_k)$, \textit{i.e.,} $P(t_k)=\cup_i S_i(t_k)$,
	\item $\Theta$: the set which collects all the operator nodes of $\mathcal{T}_{\varphi}(t_k)$, which is time invariant.
\end{itemize}
For a node $N_i(t_k)\in P(t_k)\cup \Theta$, define
\begin{itemize}
	\item $\text{CH}(N_i(t_k))$: the set of children of node $N_i(t_k)$,
	\item $\text{PA}(N_i(t_k))$: the set of parents of node $N_i(t_k)$,
	\item $\mathsf{Post}(N_i(t_k)):=\text{CH}(\text{CH}(N_i(t_k)))$,
	\item $\mathsf{Pre}(N_i(t_k)):=\text{PA}(\text{PA}(N_i(t_k)))$.
\end{itemize}

Given a state-time pair $(x_k, t_k)$, define $L: \mathbb{R}^n\times \mathbb{R}_{\ge 0}\to 2^{P(t_k)}$ as the labelling function, given by
\begin{equation}\label{labelling}
	\begin{aligned}
		&\hspace{-0cm}L(x_k, t_k)=\{S_i(t_k)\in P(t_k): x_k\in S_i(t_k), t_k\le t_h(S_i(t_k))\},
	\end{aligned}
\end{equation}
which maps $(x_k, t_k)$ to a subset of $P(t_k)$. Moreover, define the function $B: \mathbb{R}^n\times \mathbb{R}_{\ge 0}\to 2^{P(t_k)}$, which maps $(x_k, t_k)$ to a set of valid set nodes in $P(t_k)$.  The function $L(x_k, t_k)$ computes the subset of set nodes of $P(t_k)$ that contains $x_k$ at time $t_k$ (without the consideration of history trajectory) while the function $B(x_k, t_k)$ is further introduced to capture the fact that given the history trajectory, not all set nodes in $L(x_k, t_k)$ are valid at time $t_k$. A rule for determining $B(x_k, t_k)$ given $L(x_k, t_k)$ is detailed in Algorithm 7 in the next subsection.

\subsection{Online control synthesis}
In the following, we will first present the online control synthesis algorithm (and its sub-algorithms), and then an example is given to further explain how each sub-algorithm works.

\begin{algorithm}
	\caption{\textit{onlineControlSynthesis}}
	\begin{algorithmic}[1]
		\Require The tTLT $\mathcal{T}_\varphi$ and $(x_0, t_0)$.
		\Ensure ${\rm NExis}$ or $(\bm{\nu},\bm{x})$ with $\bm{\nu}=\nu_0\nu_1\ldots \nu_k\ldots$ and $\bm{x}=x_0x_1\ldots x_k\ldots$.
		\State $(t_a, t_h, \texttt{Post}(B(x_{-1}, t_{-1})))\leftarrow \textit{initialization}(\mathcal{T}_\varphi)$,
		\State $B(x_k, t_k)\leftarrow \textit{trackingSetNode}(\texttt{Post}(B(x_{k-1}, t_{k-1})))$,
		\For {each $S_i(t_k)\in B(x_k, t_k)$,}
		\If {$t_a(S_i(t_{k}))=\bowtie$,}
		\State $t_a(\mathbb{X}_i) \leftarrow t_k$,
		\EndIf
		\EndFor
		\State $\mathcal{T}_\varphi(t_{k+1})\leftarrow \textit{updatetTLT}(\mathcal{T}_\varphi(t_k),$ $t_a, B(x_k, t_k))$,
		\State $\mathcal{T}_u(t_k) \leftarrow \textit{buildControlTree}(\mathcal{T}_\varphi(t_{k}), B(x_k, t_k), \mathcal{T}_\varphi(t_{k+1}))$,
		\State $\mathcal{T}_u^c(t_k) \leftarrow \textit{Compression}(\mathcal{T}_u(t_k))$,
		\State $\mathbb{U}(x_k, t_k)\leftarrow \textit{Backtracking*}(\mathcal{T}_u^c)$,
		\If {$\mathbb{U}(x_k, t_k)= \emptyset$,}
		\State stop and return ${\rm NExis}$,
		\Else
		\State choose $\nu_k\in \mathbb{U}(x_k, t_k)$,
		\State implement $\nu_k$ and measure $x_{k+1}$,
		\State $\texttt{Post}(B(x_k, t_k))\hspace{-0.1cm} \leftarrow \hspace{-0.1cm}\textit{postSet}(B(x_k, t_k), t_a, \mathcal{T}_\varphi(t_{k+1}))$,
		\State update $k=k+1$ and go to line 2.
		\EndIf
	\end{algorithmic}
\end{algorithm}

\begin{algorithm}
	\caption{\textit{initialization}}
	\begin{algorithmic}[1]
		\Require The tTLT $\mathcal{T}_\varphi$.
		\Ensure $t_a, t_h, \texttt{Post}(B(x_{-1}, t_{-1}))$.
		\State $t_a(\mathbb{X}_{\text{root}}^\varphi) \leftarrow t_0, t_h(\mathbb{X}_{\text{root}}^\varphi) \leftarrow t_0+|\text{CH}(\mathbb{X}_{\text{root}}^\varphi)|$,
		\For {each non-root and non-leaf tube node $\mathbb{X}_i$ through a top-down traversal,}
		\State {$t_a(\mathbb{X}_i)\leftarrow \bowtie, t_h(\mathbb{X}_i)\leftarrow t_h(\mathsf{Pre}(\mathbb{X}_i)+|\text{CH}(\mathbb{X}_i)|$},
		\EndFor
		\For {each leaf node $\mathbb{X}_i$,}
		\State {$t_a(\mathbb{X}_i)\leftarrow \bowtie, t_h(\mathbb{X}_i)\leftarrow \infty,$}
		\EndFor
		\State $\texttt{Post}(B(x_{-1}, t_{-1}))\leftarrow \mathbb{X}_{\text{root}}^\varphi(t_0)$,
		\For {each $\mathbb{X}_j$ that is reachable from $\mathbb{X}_{\text{root}}^\varphi$ by a Boolean segment (see Definition \ref{Booleanfragement}),}
		\State $\texttt{Post}(B(x_{-1}, t_{-1}))\leftarrow \texttt{Post}(B(x_{-1}, t_{-1}))\cup \mathbb{X}_j(t_0)$,
		\State $t_a(\mathbb{X}_j)\leftarrow t_0$,
		\EndFor
	\end{algorithmic}
\end{algorithm}

\begin{algorithm}
	\caption{\textit{trackingSetNode}}
	\begin{algorithmic}[1]
		\Require $\texttt{Post}(B(x_{k-1}, t_{k-1}))$.
		\Ensure $B(x_{k}, t_{k})$.
		\State Compute $L(x_k, t_k)$ according to (\ref{labelling}),
		\State $B(x_k, t_k) \leftarrow L(x_k, t_k)\cap \texttt{Post}(B(x_{k-1}, t_{k-1}))$,
		\For {each $S_i(t_k)\in B(x_k, t_k)$},
		\If {$\exists S_j(t_k)\in B(x_k, t_k)$ s.t. $S_j(t_k)=\mathsf{Post}(S_i(t_k))$,}
		\State $B(x_k, t_k) \leftarrow B(x_k, t_k)\setminus S_i(t_k)$,
		\EndIf
		\EndFor
	\end{algorithmic}
\end{algorithm}

The online control synthesis algorithm is outlined in Algorithm 5. Before implementation, an initialization process (line 1) is required, which is outlined in Algorithm 6. Here, $t_a$ and $t_h$ are two functions that map each tube node $\mathbb{X}_i$ to its activation time and time horizon, respectively. If $t_a(\mathbb{X}_i)$ or $t_h(\mathbb{X}_i)$ is unknown for $\mathbb{X}_i$, its value will be set as $\bowtie$. Then, at each time instant $t_k$, a feasible control set $\mathbb{U}(x_k, t_k)$ is synthesized (lines 2-11). This process contains the following steps: 1) find the subset of set nodes in $P(t_k)$ that are valid at time $t_k$, \textit{i.e.,} $B(x_k, t_k)$, via Algorithm 7 (line 2); 2) determine the activation time of $\mathbb{X}_i$, whose corresponding set node $S_i(t_k)\in B(x_k, t_k)$ (if $t_a(\mathbb{X}_i)$ is unknown, \textit{i.e.,} being visited for the first time, it is set as $t_k$; otherwise, \textit{i.e.,} being visited before, it is unchanged) (lines 3-7); 3) calculate $\mathcal{T}_\varphi(t_{k+1})$ via Algorithm 8 (line 8); 4) build a control tree $\mathcal{T}_u(t_k)$ (Definition \ref{controltree}) via Algorithm 9 (line 9), compress it via Algorithm 3 (line 10), and then the feasible control set $\mathbb{U}(x_k, t_k)$ is given by backtracking the compressed control tree $\mathcal{T}_u^c(t_k)$ via Algorithm 10 (line 11).
If the obtained feasible control set $\mathbb{U}(x_k, t_k)=\emptyset$, the control synthesis process stops and returns ${\rm NExis}$ (lines 12-13); otherwise, the
control input $\nu_k$ can be chosen as any element of $\mathbb{U}(x_k, t_k)$ (one example is to choose $\nu_k$ as $\min_{\nu_k\in \mathbb{U}(x_k, t_k)} \{ \|\nu_k\|\}$) (line 15). Then, we implement the chosen $\nu_k$, measure $x_{k+1}$ (line 16), and finally compute the subset of set nodes that are possibly
available at the next time instant $t_{k+1}$, \textit{i.e.,} $\texttt{Post}(B(x_k, t_k))$, via Algorithm 11 (line 17).

\begin{algorithm}
	\caption{\textit{updatetTLT}}
	\begin{algorithmic}[1]
		\Require $\mathcal{T}_\varphi(t_{k})$, $t_a$ and $B(x_k, t_k)$.
		\Ensure $\mathcal{T}_\varphi(t_{k+1})$.
		\For {each set node $S_i(t_{k})$ of $\mathcal{T}_\varphi(t_{k})$,}
		\If {$S_i(t_{k})\in B(x_k, t_k) \wedge t_a(S_i(t_{k}))+|\text{CH}(S_i(t_{k}))|\ge t_{k+1}$,}
		\State $S_i(t_{k+1})\leftarrow \mathbb{X}_i(t_{k+1}-t_a(S_i(t_k)))$,
		\Else
		\State $S_i(t_{k+1})\leftarrow S_i(t_{k})$,
		\EndIf
		\EndFor
	\end{algorithmic}
\end{algorithm}

We further detail the Algorithms 6-11 in the following.
\begin{itemize}
	\item Algorithm 6 calculates the functions $t_a$ and $t_h$ (lines 1-7) and $\texttt{Post}(B(x_{-1}, t_{-1}))$ (lines 8-12).
	\item Algorithm 7 outlines the procedure of finding the subset of set nodes in $P(t_k)$ that are valid at time $t_k$, \textit{i.e.,} $B(x_k, t_k)$. This is the most important step of the control synthesis, and it relates to Algorithm 11 \textit{postSet}. Firstly, one needs to compute the subset of set nodes of $P(t_k)$ that contains $x_k$ at time $t_k$, \textit{i.e.,} $L(x_k, t_k)$ (line 1). Then, one has from Definition \ref{Def:PathSaf} that if a trajectory $\bm{x}$ satisfies one complete path of the tTLT, it must i) visit each tube node of the complete path sequentially and ii) stay in each tube node for sufficiently long time steps (Remark \ref{remark4.1}). Based on these two requirements, Algorithm 11 is designed to predict the subset of set nodes that are possibly available at the next time instant, \textit{i.e.,} $\texttt{Post}(B(x_{k-1}, t_{k-1}))$. $B(x_k, t_k)$ must belong to $L(x_k, t_k)$ and $\texttt{Post}(B(x_{k-1}, t_{k-1}))$ at the same time. Therefore, we let $B(x_k, t_k) \leftarrow L(x_k, t_k)\cap \texttt{Post}(B(x_{k-1}, t_{k-1}))$ (line 2). The rest of Algorithm 7 (lines 3-7) is to guarantee that $B(x_k, t_k)$ contains at most one set node for each complete path of $\mathcal{T}_{\varphi}(t_k)$.
	\item Algorithm 8 outlines the procedure of calculating $\mathcal{T}_\varphi(t_{k+1})$, given $\mathcal{T}_\varphi(t_{k})$, $t_a$ and $B(x_k, t_k)$. It is designed based on (\ref{sitk}).
	\item Algorithm 9 outlines the procedure of building a control tree $\mathcal{T}_u(t_{k})$, which is then used for control set synthesis. It is initialized as $\mathcal{T}_\varphi(t_{k})$ (line 1). Then, for those set nodes $S_i(t_k)$ that belongs to $B(x_k, t_k)$, it is replaced with the feasible control set (lines 2-8), otherwise, it is replaced with $\emptyset$ (lines 9-11).
	\item Algorithm 10 is similar to Algorithm 4, which outlines the procedure of backtracking a compressed tree.
	\item Algorithm 11 outlines the procedure of finding the subset of set nodes that are possibly available at the next time instant $t_{k+1}$ given $B(x_k, t_k)$, $t_a$ and $\mathcal{T}_\varphi(t_{k+1})$. It is designed based on Definition \ref{Def:PathSaf}, where the three cases (lines 4-8, 9-12, 13-16) correspond to items i)-iii) of Definition \ref{Def:PathSaf}, respectively. It guarantees that the resulting trajectory visits each tube node of $\mathcal{T}_\varphi$ sequentially and stays in each tube node for sufficiently long time steps (as we discussed in Algorithm 7).
\end{itemize}

\begin{algorithm}
	\caption{\textit{buildControlTree}}
	\begin{algorithmic}[1]
		\Require $\mathcal{T}_\varphi(t_{k})$, $B(x_k, t_k)$, and $\mathcal{T}_\varphi(t_{k+1})$.
		\Ensure A control tree $\mathcal{T}_u(t_k)$.
		\State Initialize $\mathcal{T}_u(t_k)$ as $\mathcal{T}_\varphi(t_{k})$,
		\For {each $S_i(t_k)\in B(x_k, t_k)$}
		\If {$S_i(t_k)$ is a leaf node},
		\State $S_i(t_k)\leftarrow \mathbb{U}(S_i(t_k)):=U$,
		\Else
		\State $S_i(t_k)\leftarrow\mathbb{U}(S_i(t_k)):=\{u_k\in U: f_k(x_k, u_k, w_k)\in S_i(t_{k+1}), \forall w_k\in W\}$,
		\EndIf
		\EndFor
		\For {each $S_i(t_k)\notin B(x_k, t_k)$}
		\State $S_i(t_k)\leftarrow \emptyset$,
		\EndFor
	\end{algorithmic}
\end{algorithm}

\begin{algorithm}
	\caption{\textit{Backtracking*}}
	\begin{algorithmic}[1]
		\Require a compressed tree $\mathcal{T}_u^c(t_k)$.
		\Ensure the root node of $\mathcal{T}_u^c(t_k)$.
		\For {each Boolean operator node $\Theta$ of $\mathcal{T}_u^c(t_k)$ through a bottom-up traversal,}
		\If {$\Theta=\wedge$,}
		\State $\text{PA}(\Theta)\leftarrow\text{PA}(\Theta)\cup (\text{CH}_1(\Theta)\cap \text{CH}_2(\Theta))$,
		\Else
		\State $\text{PA}(\Theta)\leftarrow\text{PA}(\Theta)\cup (\text{CH}_1(\Theta)\cup \text{CH}_2(\Theta))$,
		\EndIf
		\EndFor
	\end{algorithmic}
\end{algorithm}

\begin{algorithm}
	\caption{\textit{postSet}}
	\begin{algorithmic}[1]
		\Require $B(x_{k}, t_{k}), t_a$ and $\mathcal{T}_\varphi(t_{k+1})$.
		\Ensure $\texttt{Post}(B(x_{k}, t_{k}))$.
		\State Initialize $\texttt{Post}(S_i(t_k))=\emptyset, \forall S_i(t_k)\in B(x_{k}, t_{k})$.
		\For {each $S_i(t_k)\in B(x_{k}, t_{k})$,}
		\Switch{the children of $S_i(t_k)$}
		\Case{$\text{CH}(S_i(t_k))\in \{\wedge, \vee\}$,}
		\State $\texttt{Post}(S_i(t_{k}))\leftarrow S_i(t_{k+1})$,
		\For {each $S_j(t_k)$ that is reachable from $S_i(t_k)$ by a Boolean segment,}
		\State $\texttt{Post}(S_i(t_{k}))\leftarrow \texttt{Post}(S_i(t_{k}))\cup S_j(t_{k+1})$,
		\EndFor
		\EndCase
		\Case{$\text{CH}(S_i(t_k))\in \{\mathsf{U}_{[a, b]}, \mathsf{F}_{[a, b]}\}$,}
		\If{$t_k>t_a(\mathsf{Pre}(S_i(t_k))+a\big\}$,}
		\State $\texttt{Post}(S_i(t_{k}))\leftarrow S_i(t_{k+1}) \cup\mathsf{Post}(S_i(t_{k+1}))$,
		\EndIf
		\EndCase
		\Case{$\text{CH}(S_i(t_k))\in \{\mathsf{G}_{[a, b]}\}$,}
		\If{$t_k>t_a(\mathsf{Pre}(S_i(t_k))+b\big\}$,}
		\State $\texttt{Post}(S_i(t_{k}))\leftarrow S_i(t_{k+1}) \cup\mathsf{Post}(S_i(t_{k+1}))$,
		\EndIf
		\EndCase
		\EndSwitch
		\EndFor
	\end{algorithmic}
\end{algorithm}

Next, an example is given to illustrate one iteration of the control synthesis algorithm (Algorithm 5).
\begin{example}\label{example2}
	Consider the single-integrator control system $\dot x=u+w$ with a sampling period of one second.  The corresponding discrete-time system is given by
	\begin{equation*}
		x_{k+1}=x_k+u_k+w_k,
	\end{equation*}
	where $x_k\in \mathbb{R}^2, u_k\in U:=\{u: ||u||\le 1\}\subset \mathbb{R}^2, w_k\in W :=\{w: ||w||\le 0.1\}\subset \mathbb{R}^2, \forall k\in \mathbb{N}$. The task specification $\varphi$ is given in Example \ref{example1}, \textit{i.e.,} $\varphi=\mathsf{F}_{[a_1, b_1]}\mathsf{G}_{[a_2, b_2]}\mu_1 \wedge \mu_2 \mathsf{U}_{[a_3, b_3]}\mu_3$, where $[a_1, b_1]=[5, 10]$, $[a_2, b_2]=[0, 10]$, $[a_3, b_3]=[0, 8]$, $g_{\mu_1}(x)=1-\|x\|$, $g_{\mu_2}(x)=5-\|x-[4, 4]^T\|$, and $g_{\mu_3}(x)=1-\|x-[3, 5]^T\|$. Then, one has
	\begin{eqnarray*}
		&&\mathbb{S}_{\mu_1}=\{x_0: \|x_0\|\le 1\}, \\
		&& \mathbb{S}_{\mu_2}=\{x_0: \|x_0-[4, 4]^T\|\le 5\},\\
		&&\mathbb{S}_{\mu_3}=\{x_0: \|x_0-[3, 5]^T\|\le 1\}.
	\end{eqnarray*}
	The tTLT that corresponds to $\varphi$ is plotted in Figure \ref{Fig:tTLT}. Using Definitions \ref{Def:maxreachset} and \ref{Def:minreachset}, one can calculate that
	\begin{eqnarray*}
		&&\hspace{-0.5cm}\mathbb{X}_4(t_k)=\{x_k: \|x_k\|\le 0.9\}, \\
		&& \hspace{-0.5cm}\mathbb{X}_3(t_k)=\{x_k: \|x_k-[3, 5]^T\|\le 8.1-k \\
		&& \hspace{3.5cm}\wedge \|x_k-[4, 4]^T\|\le 5\}, \\
		&&\hspace{-0.5cm}\mathbb{X}_2(t_k)=\{x_k: \|x_k\|\le 9.9-k\}, \\
		&&\hspace{-0.5cm}\mathbb{X}_1(t_k)=\mathbb{X}_2(t_k)\cap \mathbb{X}_3(t_k).
	\end{eqnarray*}
	The initial state $x_0=[0.5, 0.8]^T$, for which $x_0\in \mathbb{X}_{\text{root}}^\varphi(t_0)$. Firstly, an initialization process is required, and one can get from Algorithm 6 that
	\begin{eqnarray*}
		&&t_h(\mathbb{X}_1)=0, t_h(\mathbb{X}_2)=10, t_h(\mathbb{X}_3)=8, \\ &&t_h(\mathbb{X}_4)=20,
		t_h(\mathbb{S}_{\mu_1})=\infty, t_h(\mathbb{S}_{\mu_3})=\infty,
	\end{eqnarray*}
	and
	\begin{equation*}
		\texttt{Post}(B(x_{-1}, t_{-1}))=\{\mathbb{X}_1(t_0), \mathbb{X}_2(t_0), \mathbb{X}_3(t_0)\}.
	\end{equation*}
	
	Now, let us see how the feasible control set $\mathbb{U}(x_0, t_0)$ is synthesized at time instant $t_0$.
	
	\begin{figure*}
		\centering
		\includegraphics[width=0.7\textwidth]{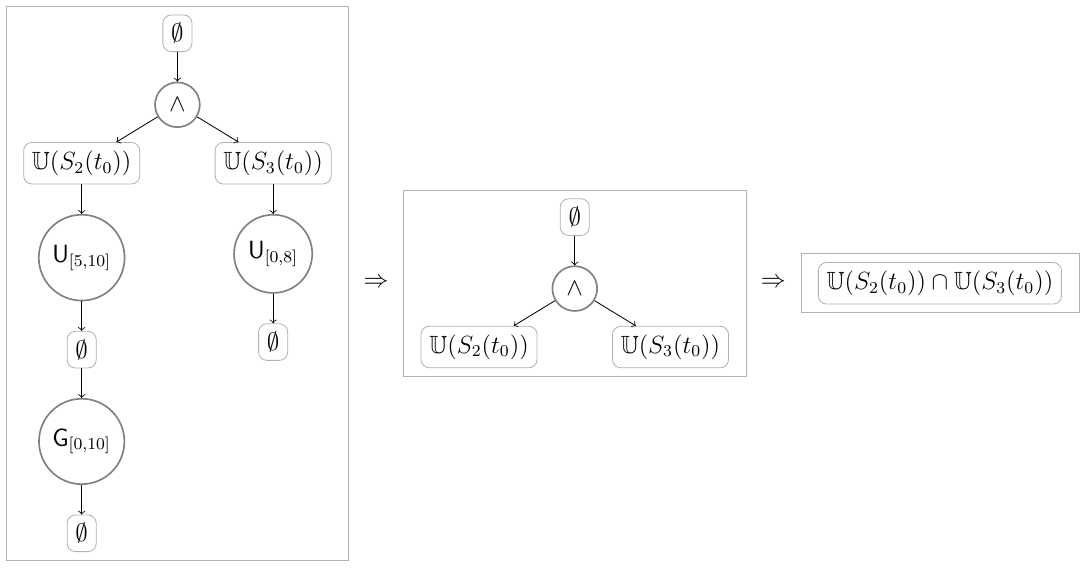}
		\caption{Left: $\mathcal{T}_u(t_0)$, Middle: $\mathcal{T}_u^c(t_0)$, Right: root node of $\mathcal{T}_u^c(t_0)$ after implementing Algorithm 10, where $\mathbb{U}(S_2(t_0))=U=\{u: ||u||\le 1\}, \mathbb{U}(S_3(t_0))=U\cap \{u: \|u-[3.4, 3.1]^T\|\le 5\}$.}
		\label{Fig:control_ex1}
	\end{figure*}
	
	1) Find $B(x_0, t_0)$ via Algorithm 7. Firstly, $L(x_0, t_0)$ is computed according to (\ref{labelling}),
	\begin{equation*}
		L(x_0, t_0)=\{\mathbb{X}_1(t_0), \mathbb{X}_2(t_0), \mathbb{X}_3(t_0), \mathbb{X}_4(t_0), \mathbb{S}_{\mu_1}\}.
	\end{equation*}
	Then, after running lines 2-7, one has
	\begin{equation*}
		B(x_0, t_0)=\{S_2(t_0), S_3(t_0)\}.
	\end{equation*}
	
	2) Determine the activation time. Initially, both $t_a(\mathbb{X}_2)$ and $t_a(\mathbb{X}_3)$ are unknown, therefore, $t_a(\mathbb{X}_2)=t_a(\mathbb{X}_3)=t_0$.
	
	3) Update the TLT (thus obtain $\mathcal{T}_\varphi(t_1)$) via Algorithm 8. The output $\mathcal{T}_\varphi(t_1)$ is given by
	\begin{eqnarray*}
		&&\hspace{-0.5cm}S_1(t_1)=\mathbb{X}_1(t_0), S_2(t_1)=\mathbb{X}_2(t_1), \\
		&&\hspace{-0.5cm}S_3(t_1)=\mathbb{X}_3(t_1), S_4(t_1)=\mathbb{X}_4(t_0),
	\end{eqnarray*}
	and the leaf nodes $\mathbb{S}_{\mu_1}$ and $\mathbb{S}_{\mu_3}$ are unchanged.
	
	4)  Build the control tree $\mathcal{T}_u(t_0)$, compress it to obtain $\mathcal{T}_u^c(t_0)$, and then get $\mathbb{U}(x_0, t_0)$. This process is illustrated in Figure \ref{Fig:control_ex1}, and $\mathbb{U}(x_0, t_0)=\mathbb{U}(S_2(t_0))\cap \mathbb{U}(S_3(t_0))$.
	
	Since $\mathbb{U}(x_0, t_0)\neq \emptyset$, the online control synthesis continues, and we can further compute
	$\texttt{Post}(B(x_{0}, t_{0}))$ via Algorithm 11, which gives
	\begin{equation*}
		\texttt{Post}(B(x_{0}, t_{0}))=\{S_2(t_1), S_3(t_1), \mathbb{S}_{\mu_3}\}.
	\end{equation*}
\end{example}

The following theorem and corollary show the applicability and correctness of Algorithm 5.
\begin{theorem}\label{The:RecFea}
	Consider the uncertain system (\ref{x0}) with initial state $x_0$ and an STL formula $\varphi$ in (\ref{Def:PNF}). Assume that $\varphi$ is robustly satisfiable for (\ref{x0}) and $x_0\in \mathcal{T}_{\text{root}}^\varphi(t_0)$. Then, by implementing the online control synthesis algorithm (Algorithm 5), one can guarantee that
	\begin{itemize}
		\item[(i)]  the control set $\mathbb{U}(x_k, t_k)$ is nonempty for all $k\in \mathbb{N}$;
		\item[(ii)]  the resulting trajectory $\bm{x}\vDash\varphi$.
	\end{itemize}
\end{theorem}
\begin{proof}
	The proof follows from the construction of tTLT and Algorithms 5-11. The existence of a controller $\nu_k$ at each time step $t_k$, is guaranteed by the definition of maximal and minimal reachable sets (Definitions \ref{Def:maxreachset} and \ref{Def:minreachset}), and the construction of tTLT (Propoition \ref{prop1}, Theorem \ref{thm1} and Algorithm 1). Moreover, the design of Algorithms 5-11 guarantees that the resulting trajectory $\bm{x}$ satisfies the tTLT $\mathcal{T}_\varphi$, \textit{i.e.,} $\bm{x}\vDash \mathcal{T}_\varphi$, which implies $\bm{x}\vDash \varphi$ as proven in Theorem \ref{The:SLTtLTL}.
 \qed
\end{proof}

\begin{remark}
	The tTLT construction relies on the computation of backward reachable tubes. Over the past decade, new approaches (e.g., decomposition-based approach \citep{chen2018decomposition} and learning-based approaches \citep{allen2014machine,bansal2020deepreach}) and software tools (e.g., Hamilton-Jacobi Toolbox \citep{mitchell2005toolbox} and CORA Toolbox \citep{althoff2015introduction}), have been developed for improving the efficiency of computing backward reachable tubes. Moreover, we remark that the computation of reachable tubes in our work for constructing of the tTLT can be performed \emph{offine}, which may mitigate the online computational burden. On the other hand, although the exact computation of backward reachable sets/tubes is in general nontrivial for high-dimensional nonlinear systems, efficient algorithms exist for linear systems with polygonal input and disturbance sets \citep{kurzhanski2014}.
\end{remark}

\begin{remark}
	The online control synthesis algorithm (Algorithm 5) contains 7 sub-algorithms, \textit{i.e.,} Algorithm 3 and Algorithms 6-11. The computational complexity is determined by Algorithm 9, in which one-step feasible control sets need to be computed. The computational complexity of Algorithms 3, 6, 7, 8, 10, 11 is $\mathcal{O}(1)$. Note that in Algorithm 8, the computation of reachable sets, which is required for set node update, is done offline when constructing the tTLT.
\end{remark}

\begin{remark}
	Different from the mixed-integer programming formulation for STL control synthesis \citep{raman2015,raman2014}, where an entire control policy has to be synthesized at each time step, the control synthesis in our work is reactive in the sense that only the control input at the current time step is generated at each time step. 
\end{remark}

\section{Case Studies}
In this section, two  examples illustrating the theoretical results are provided.  We first perform a numerical simulation for car overtaking and then apply our algorithms to a  car parking scenario.

\subsection{Car overtaking example}

\begin{figure*}
	\centering
	\subfigure{
		\includegraphics[width=0.7\textwidth]{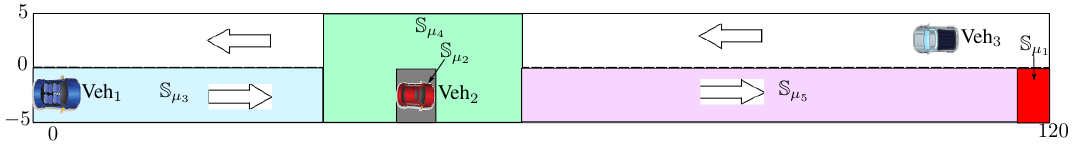}}
	\caption{\footnotesize Scenario illustration: an automated vehicle plans to reach a target set $\mathbb{S}_{\mu_1}$ while overtaking a broken vehicle $\text{Veh}_2$ in front of it in the same lane and avoiding $\text{Veh}_3$ moving in an opposite direction in the other lane. }
	\label{Fig:examplescenario}
\end{figure*}

\begin{figure}
	\centering
	\subfigure{ \includegraphics[width=0.35\textwidth]{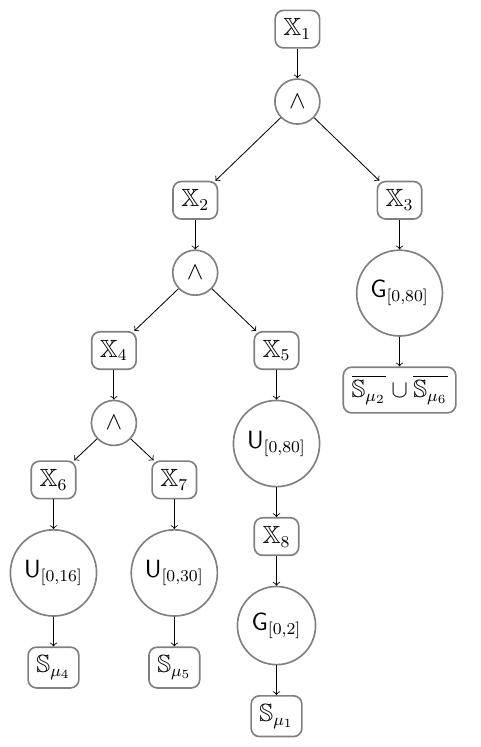}}
	\caption{\footnotesize The constructed tTLT $\mathcal{T}_{\varphi_{\rm fast\_{overtake}}}$.}
	\label{Fig:overtakingTLT}
\end{figure}

We first consider a car overtaking example. This example will specify an overtaking task as an STL formula and then show how to synthesize overtaking controller with safety guarantee.

As shown in Figure~\ref{Fig:examplescenario}, we consider a scenario where an automated vehicle $\text{Veh}_1$ plans to move to a target set $\mathbb{S}_{\mu_1}$ within $80$ seconds. Since there is a broken vehicle $\text{Veh}_2$ in front of $\text{Veh}_1$ and there is another vehicle $\text{Veh}_3$ that moves in an opposite direction in the other lane, $\text{Veh}_1$ must overtake $\text{Veh}_2$ for reaching $\mathbb{S}_{\mu_1}$  and avoid $\text{Veh}_3$ for safety.

We describe the dynamics of the vehicle $\text{Veh}_1$ as in \citet{murgovski2015}:
\begin{eqnarray*}
	x_{k+1}=\underbrace{\left[\begin{array}{ccccccc}
			1 & 0 & \delta  \\
			0 & 1 & 0 \\
			0 & 0 & 1
		\end{array}\right]}_{A}x_{k}+ \underbrace{\left[\begin{array}{ccccccc}
			0 & 0 \\
			\delta & 0 \\
			0  &  \delta
		\end{array}\right]}_{B}u_{k}+w_k,
\end{eqnarray*}
where $x_{k}=[p^x(k), p^y(k), v^x(k)]^T$, $u_{k}=[v^y(k), a^x(k)]^T$, and $\delta$ is the sampling period.
The working space is $X=\{z\in \mathbb{R}^3\mid [0, -5, -3]^T \leq z\leq [120, 5, 3]^T\}$, the control  constraint set is $U=\{z\in \mathbb{R}^2\mid [-1, -1]^T \leq z\leq [1, 1]^T \}$, the disturbance set is  $W=\{z\in \mathbb{R}^3\mid [-0.05, -0.05, -0.05]^T \leq z\leq [0.05, 0.05, 0.05]^T\}$, and the target region is  $\mathbb{S}_{\mu_1}=\{z\in \mathbb{R}^2\mid [115, -5,0.5]^T \leq z\leq [120, 0, 0.5]^T\}$.

We use $\mathbb{S}_{\mu_2}=\{z\in \mathbb{R}^3\mid [45, -5, -\infty]^T \leq z\leq [50, 0, \infty]^T\}$ to denote the state set that contains the occupancy of $\text{Veh}_2$.
We describe  the dynamics  of the vehicle $\text{Veh}_3$ as
\begin{eqnarray*}
	\bar{x}_{k+1}=\underbrace{\left[\begin{array}{ccccccc}
			1 & 0 \\
			0 & 1
		\end{array}\right]}_{\bar{A}}x_{k}+ \underbrace{\left[\begin{array}{ccccccc}
			\delta & 0 \\
			0  &  \delta
		\end{array}\right]}_{\bar{B}}\bar{u}_{k},
\end{eqnarray*}
where  $x_{k}=[\bar{p}^x(k), \bar{p}^y(k)]^T$, $\bar{u}_{k}=[\bar{v}^x(k), \bar{v}^y(k)]^T$,
We assume that it moves at a constant velocity $\bar{u}_{k}=[\bar{v}^x, 0]^T$. The initial state of $\mathcal{V}_3$ is $\bar{x}_{0}=[\bar{p}^x_{ini}, 2.5]^T$.  Then, we have that its position of $x$-axis is $\bar{p}^x_k=\bar{p}^x_{ini}+\delta\times(k-1)\times\bar{v}^x$.

\begin{figure*}
	\centering
	\subfigure{
		\includegraphics[width=0.8\textwidth]{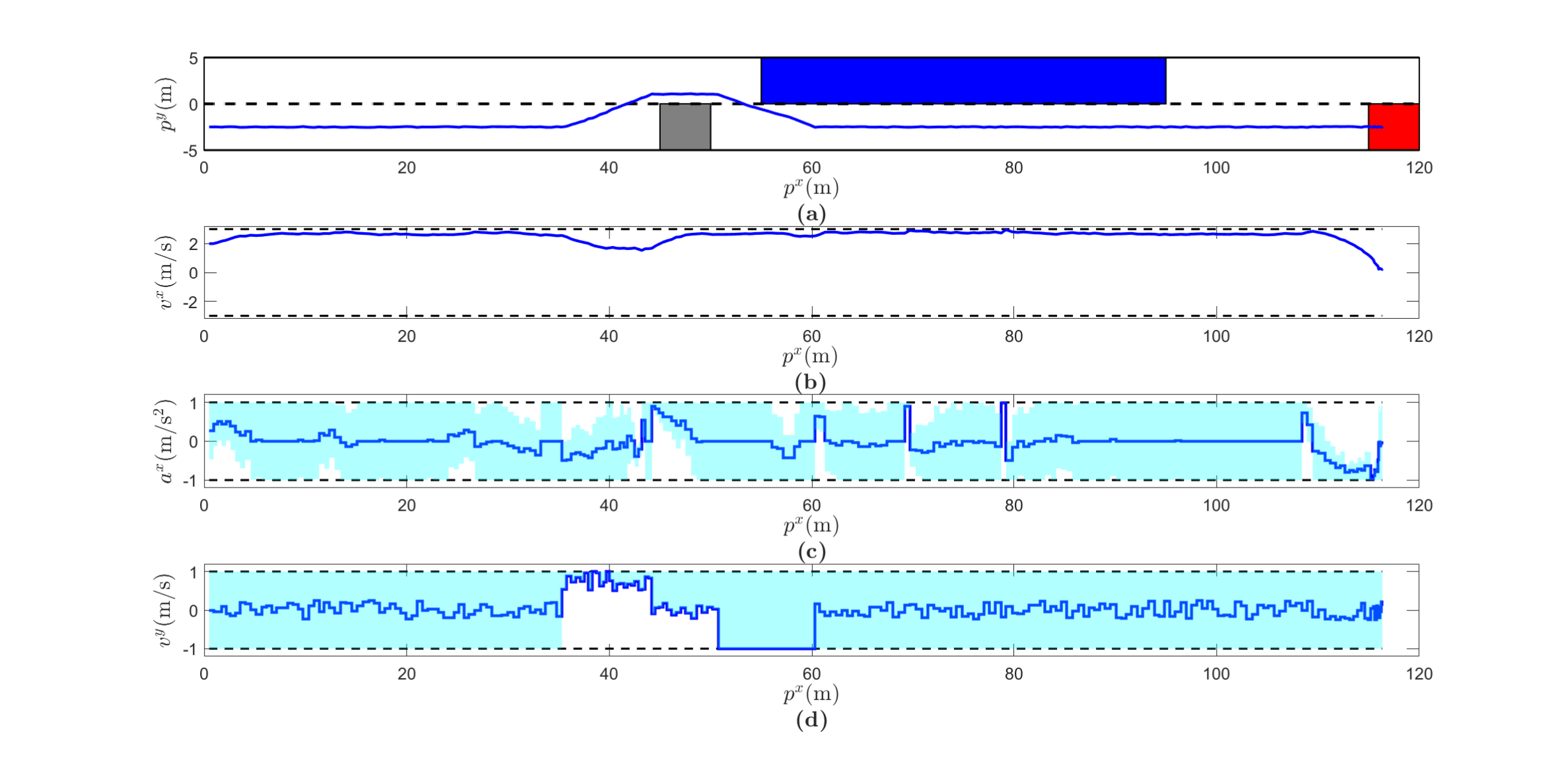}}
	\caption{\footnotesize  Trajectories for one realization of disturbance signal in the fast overtaking:  (a) position trajectory; (b) velocity trajectory of $x$-axis; (c) control trajectory of $x$-axis; (d) control trajectory of $y$-axis.}
	\label{Fig:Exam2_case1}
\end{figure*}

\begin{figure*}
	\centering
	\subfigure{
		\includegraphics[width=0.8\textwidth]{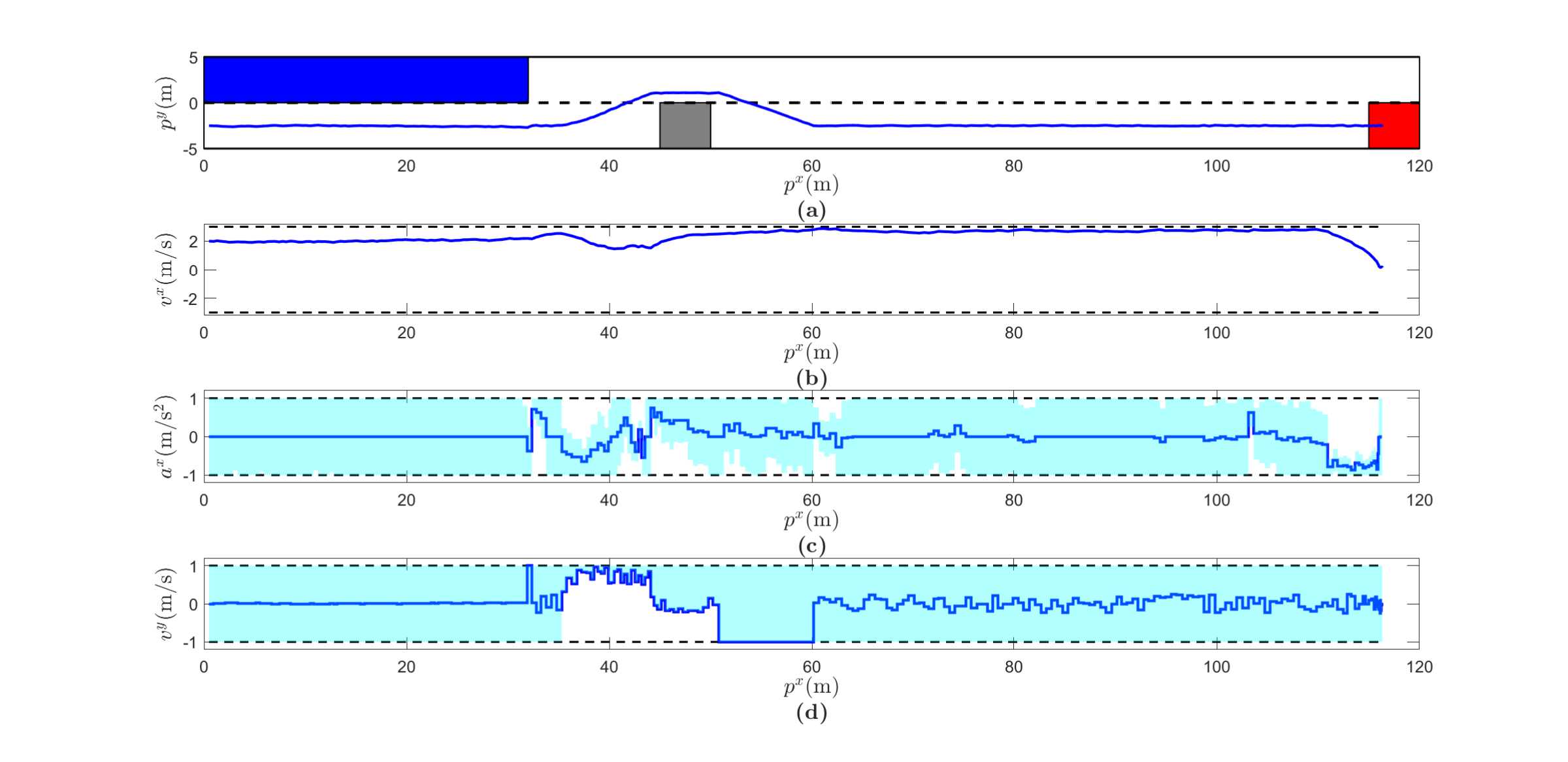}}
	\caption{\footnotesize  Trajectories for one realization of disturbance signal in the slow overtaking:  (a) position trajectory; (b) velocity trajectory of $x$-axis; (c) control trajectory of $x$-axis; (d) control trajectory of $y$-axis.}
	\label{Fig:Exam2_case2}
\end{figure*}

To formulate the overtaking task, we define the following three  sets as shown in Figure~\ref{Fig:examplescenario}: $\mathbb{S}_{\mu_3}=\{z\in \mathbb{R}^3\mid [0, -5,-3]^T \leq z\leq [35, 0,3]^T\}$, $\mathbb{S}_{\mu_4}=\{z\in \mathbb{R}^3\mid [35, -5,-3]^T \leq z\leq [60, 5,3]^T\}$, and $\mathbb{S}_{\mu_5}=\{z\in \mathbb{R}^3\mid [60, -5,-3]^T \leq z\leq [120, 0,3]^T\}$.

Let us choose the sampling period as $\delta=0.2s (\text{seconds})$.
To respect the time constraint and the input constraint for~$\text{Veh}_1$, we consider two possible solutions to the previous  reachability problem: (1) fast overtaking: overtake $\text{Veh}_2$  before $\text{Veh}_3$ passes $\text{Veh}_2$; (2) slow overtaking: wait until $\text{Veh}_3$ passes $\text{Veh}_2$ and then overtake $\text{Veh}_2$. The fast overtaking can be encoded into an STL formula:
\begin{equation}\label{type1}
	\begin{aligned}
		&\varphi_{\rm fast\_{overtake}}=\mu_3\mathsf{U}_{[0, 16]}\mu_4 \wedge 
		(\mu_3\vee \mu_4) \mathsf{U}_{[0, 30]} \mu_5 \\ & \hspace{1cm} \wedge  (\mu_3\vee \mu_4\vee \mu_5)\mathsf{U}_{[0, 80]}\mathsf{G}_{[0, 2]} \mu_1 \wedge  \mathsf{G}_{[0, 80]} \neg (\mu_2\vee\mu_6),
	\end{aligned}
\end{equation}
where $\mathbb{S}_{\mu_6}=\{z\in \mathbb{R}^6\mid [\bar{p}^x(16), 0, -\infty]^T \leq z\leq [\bar{p}^x(0), 5, \infty]^T\}$. Note that $\mathbb{S}_{\mu_6}$ denotes the reachable set for the vehicle $\text{Veh}_3$ within the time interval $[0,16]$ seconds and $16$ (that corresponds to the sampling index $k=80$) is the maximal time instant that the vehicle $\text{Veh}_1$ can reach the set $\mathbb{S}_{\mu_5}$ in the sprit of $\varphi_1$. Using Algorithm 1, one can construct the tTLT $\mathcal{T}_{\varphi_{\rm fast\_{overtake}}}$ (see Figure \ref{Fig:overtakingTLT}), where \begin{eqnarray*}
	&&\mathbb{X}_6(t_k)=\mathcal{R}^M(X, \mathbb{S}_{\mu_4}, \mathbb{S}_{\mu_3}, [0, 16], k),\\
	&&\mathbb{X}_7(t_k)=\mathcal{R}^M(X, \mathbb{S}_{\mu_5},\mathbb{S}_{\mu_3}\cup \mathbb{S}_{\mu_4}, [0, 30], k),\\
	&&\mathbb{X}_8(t_k)=\overline{\mathcal{R}^m(X, \overline{\mathbb{S}_{\mu_1}}, [0, 2], k)},\\
	&&\mathbb{X}_4(t_k)=\mathbb{X}_6(t_k)\cap \mathbb{X}_7(t_k), \\ 
	&&\mathbb{X}_5(t_k)=\mathcal{R}^M(X, \mathbb{X}_8(t_0), \mathbb{S}_{\mu_3}\cup \mathbb{S}_{\mu_4}\cap \mathbb{S}_{\mu_5}, [0, 80], k),\\
	&&\mathbb{X}_2(t_k)=\mathbb{X}_4(t_k)\cap \mathbb{X}_5(t_k),\\
	&&\mathbb{X}_3(t_k)=\overline{\mathcal{R}^m(X, \mathbb{S}_{\mu_2}\cap \mathbb{S}_{\mu_6}, [0, 80], k)},\;\text{and}\\
	&&\mathbb{X}_1(t_k)=\mathbb{X}_2(t_k)\cup \mathbb{X}_3(t_k).
\end{eqnarray*}

The slow overtaking can be encoded into an STL formula
\begin{equation}\label{type2}
	\begin{aligned}
		&\varphi_{\rm slow\_{overtake}}=\mu_3\mathsf{U}_{[16, 32]}\mu_4 \wedge 
		(\mu_3\vee \mu_4) \mathsf{U}_{[0, 45]} \mu_5 \\ & \hspace{1cm} \wedge  (\mu_3\vee \mu_4\vee \mu_5)\mathsf{U}_{[0, 80]}\mathsf{G}_{[0, 2]} \mu_1 \wedge  \mathsf{G}_{[0, 80]} \neg (\mu_2\vee\mu_7)
	\end{aligned}
\end{equation}
where $\mathbb{S}_{\mu_7}=\{z\in \mathbb{R}^2\mid [-\infty, 0,-\infty]^T \leq z\leq [\bar{p}^x(16), 5, \infty]^T\}$. Note that $\mathbb{S}_{\mu_7}$ denotes the reachable set for the vehicle Veh$_3$ within the time interval $[16, +\infty)$ and $16$ (that corresponds to the sampling index $k=80$)  is the minimal time instant that the vehicle Veh$_1$ can reach the set $\mathbb{S}_{\mu_4}$ in the sprit of $\varphi_2$. The tTLT $\mathcal{T}_{\varphi_{\rm slow\_{overtake}}}$ can be constructed similar to $\mathcal{T}_{\varphi_{\rm fast\_{overtake}}}$.

\begin{figure*}
	\centering
	\subfigure{ \includegraphics[width=0.80\textwidth]{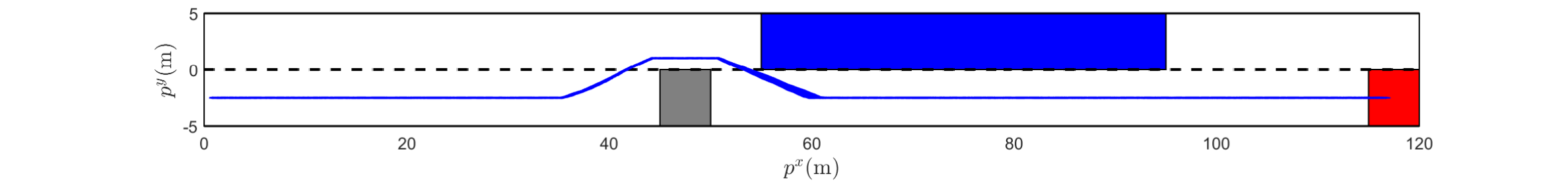}}
	\caption{\footnotesize  State trajectories for 100 realizations of disturbance signals in the fast overtaking.}
	\label{Fig:Exam2_100realization}
\end{figure*}

\begin{figure}
	\centering
	\subfigure{ \includegraphics[width=0.35\textwidth]{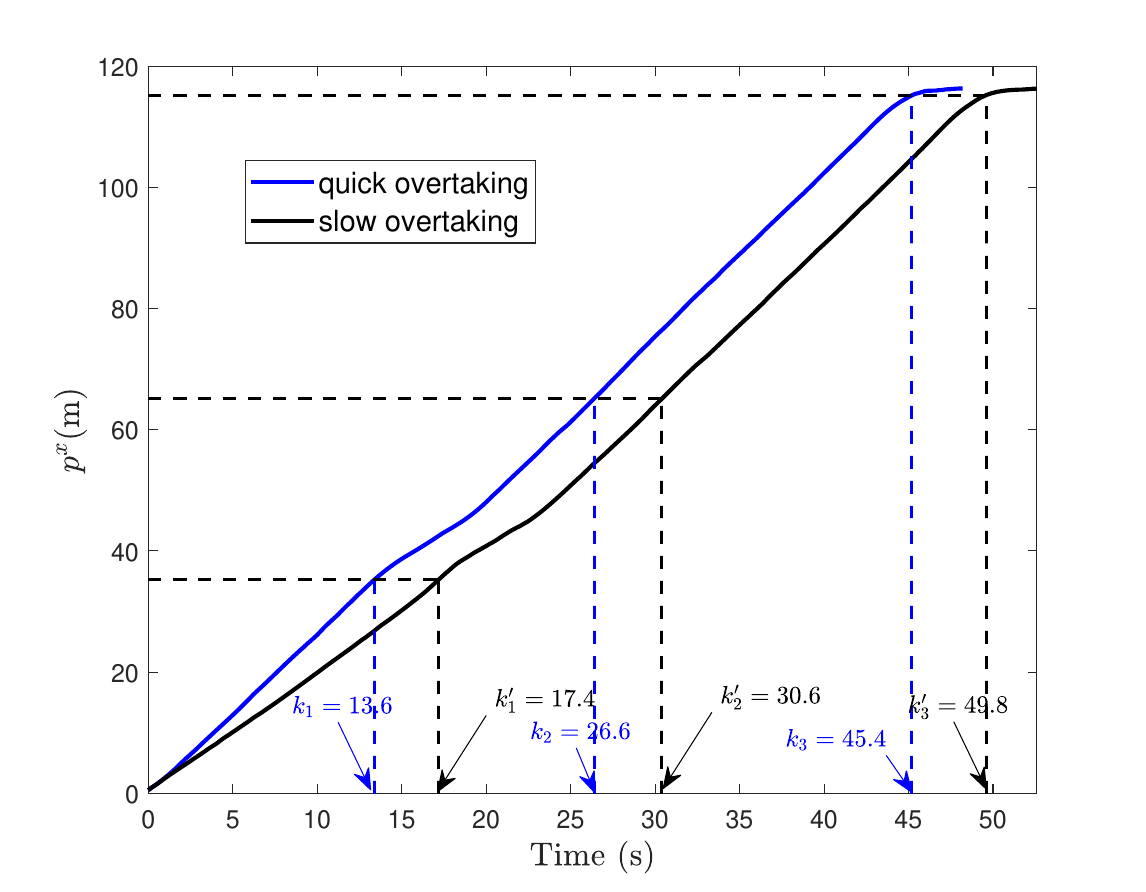}}
	\caption{\footnotesize The position trajectories of $x$-axis along the time for both type of overtaking STL tasks as defined in (\ref{type1}) and (\ref{type2}).}
	\label{Fig:positiontime}
\end{figure}

In the following, two simulation cases are considered and the online control synthesis algorithm is implemented. In the fast overtaking,  we choose the initial position $\bar{p}^x_{ini}=95$ and the moving velocity  $\bar{v}^x=-2$ for the vehicle $\text{Veh}_3$ and the initial position $x_0=[0.5,-2.5,2]^T$ for $\text{Veh}_1$. One can verify that the specification  $\varphi_{\rm slow\_{overtake}}$ is infeasible in this case. Figure~\ref{Fig:Exam2_case1} (a) shows the position trajectories, from which we can see that the whole specification is fulfilled.  The blue region denotes the set $\mathbb{S}_{\mu_6}$.
Figure~\ref{Fig:Exam2_case1} (b)  shows the velocity trajectory of $v^x$ and
Figures~\ref{Fig:Exam2_case1} (c)--(d) show the corresponding control inputs, where the dashed lines denote the control bounds.
The cyan regions represent the synthesized control
sets and the blue lines are the control trajectories.
In the slow overtaking,  we choose the initial position $\bar{p}^x_{ini}=80$ and the moving velocity  $\bar{v}^x=-3$ for the vehicle $\text{Veh}_3$ and the same initial position $x_0=[0.5,-2.5]^T$ for $\text{Veh}_1$. In this case one can verify that $\varphi_{\rm fast\_{overtake}}$ is infeasible. Figure~\ref{Fig:Exam2_case2} (a) shows the position trajectories, from which we can see that the whole specification is fulfilled.  The blue region denotes the intersection between the set $X$ and  the set $\mathbb{S}_{\mu_7}$. Figure~\ref{Fig:Exam2_case2} (b)  shows the velocity trajectory of $v^x$ and Figures~\ref{Fig:Exam2_case2} (c)--(d) show the corresponding control input trajectories of $a^x$ and $v^y$.

Although the position trajectories in the two cases are similar as shown in Figures~\ref{Fig:Exam2_case1}(a)--\ref{Fig:Exam2_case2}(a), we highlight their difference through the evolution of the position of $x$-axis along the time in Figure~\ref{Fig:positiontime}. We use $k_1$, $k_2$, and $k_3$ (or $k'_1$, $k'_2$, and $k'_3$ ) to denote the minimal time instants that $\text{Veh}_1$ reaches the sets $\mathbb{S}_{\mu_4}$, $\mathbb{S}_{\mu_5}$, and $\mathbb{S}_{\mu_1}$ in the fast overtaking (or the slow overtaking), respectively. We can see that these two position trajectories satisfy the time intervals encoded in the  $\varphi_1$ and $\varphi_2$, respectively. Furthermore, in order to show the robustness, we  run 100 realizations of the disturbance trajectories in the fast overtaking. The position trajectories for such 100 realizations are shown in Figure~\ref{Fig:Exam2_100realization}.

Finally, we report the computation time of this example, which was run in Matlab R2016a with MPT toolbox \citep{herceg2013} on a Dell laptop with Windows 7, Intel i7-6600U CPU 2.80 GHz and 16.0 GB RAM. We perform reachability analysis for constructing the tTLT offline, which takes 59.10 seconds. For online control synthesis, the minimal computation time  at a single time step over 100 realizations is 0.23 seconds, while the maximal computation time is 1.07 seconds. The average time of each time step is 0.31 seconds. We remark that the mixed-integer formulation is difficult to implement in this example. This is because the computational complexity of mixed-integer programming  grows exponentially with the horizon of the STL formula, which in this example reaches up to 400 sampling instants, much longer than the horizons considered in the simulation examples of \citet{raman2015,raman2014,sadraddini2015}.

\subsection{Car parking example}
	
	Next, we consider a car parking example. This example will specify a parking task as an STL formula and then show how our algorithms perform on real hardware. We will first perform reachability analysis for constructing the tTLT offline and then we use the tTLT to synthesize a parking controller for the Small-Vehicles-for-Autonomoy (SVEA) platform~\citep{jiang2022}. 
	
	\begin{figure}[b]
		\centering
		\includegraphics[width=0.4\textwidth]{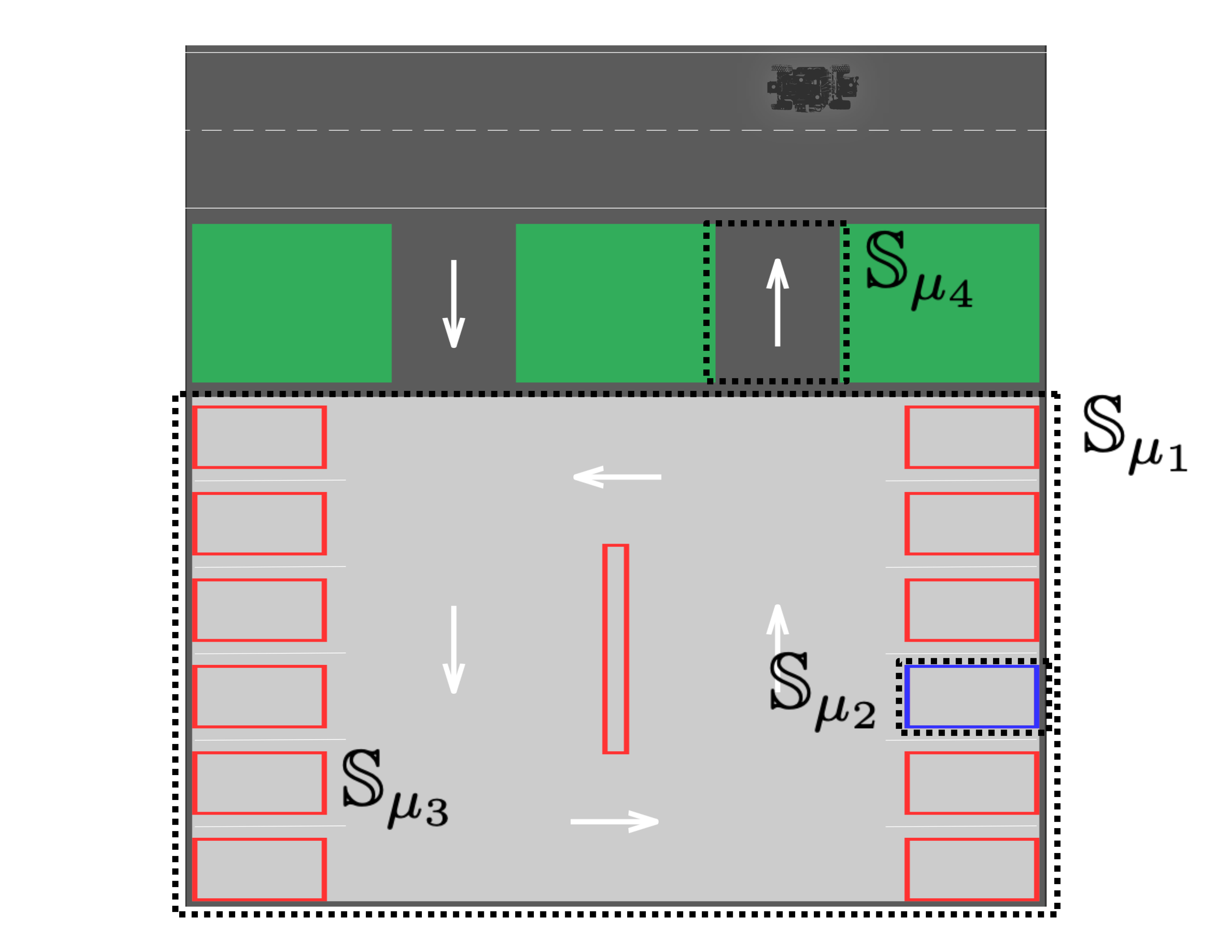}
		\caption{Scenario illustration: an automated vehicle needs to enter into the parking lot, park in the designated parking spot (blue), and leave the parking lot, while avoiding any collisions.}
		\label{Fig:parkingscenario}
	\end{figure}
	As shown in Figure~\ref{Fig:parkingscenario}, we consider a scenario where an automated vehicle must enter the parking lot $\mathbb S_{\mu_1}$, park in the designated parking spot $\mathbb S_{\mu_2}$, and leave the parking lot through the exit $\mathbb S_{\mu_4}$, where each step of the scenario has a specific deadline. Additionally, throughout the scenario, the vehicle must stay safe and avoid collisions with the parking lot walls and parked vehicles $\mathbb S_{\mu_3}$.
	
	We describe the underlying continuous dynamics of the automated vehicle as:
	\begin{equation}\label{eq:vehicle_dynamics}
		f(x, u, w) =
		\left[\begin{matrix}
			\dot{p_x}\\
			\dot{p_y}\\
			\dot{\theta}\\
			\dot{v}
		\end{matrix}\right] =
		\left[\begin{matrix}
			v \cos\theta \\
			v \sin\theta \\
			\frac{v \tan \delta}{L}\\
			a
		\end{matrix}\right] + w,
	\end{equation}
	where $x = [p_x, p_y, \theta, v]^T$ is the vehicle's $x$ position, $y$ position, heading, and velocity, respectively. $u = [\delta, a]^T$ is the vehicle's steering and acceleration inputs. The working space is $X = \{z \in \mathbb R^4 \mid [-2, -3, -\pi, -0.6]^T \leq z \leq [2, 2, \pi, 0.6]^T\}$, the control set is $U = \{z \in \mathbb R^2 \mid [-\pi/5, -0.5]^T \leq z \leq [\pi/5, 0.5]^T\}$, and the disturbance set is $W = \{z \in \mathbb R^4 \mid [-0.01, -0.01, -\pi/72, -0.01]^T \leq z \leq [0.01, 0.01, \pi/72, 0.01]^T\}$. For constructing the tTLT, we discretize~\eqref{eq:vehicle_dynamics} using a simple zero-order hold estimation. Let $\delta$ be the sampling period, then we describe the discrete dynamics of the automated vehicle as
	\begin{equation}
		x_{k+1} = x_k + f(x_k, u_k, w_k)\delta.
	\end{equation}
	For the parking task, we set $\delta=0.05$s. We define the state sets in Figure~\ref{Fig:parkingscenario} as $\mathbb S_{\mu_1} = \{z \in \mathbb R^4 \mid [-2, -3, -\pi, -0.6]^T \leq z \leq [2, 0, \pi, 0.6]^T\}$, 
	$\mathbb S_{\mu_2} = \{z \in \mathbb R^4 \mid [1.3, -2, -\pi, -0.6]^T \leq z \leq [2, -1.5, \pi, 0.6]^T\}$,
	$\mathbb S_{\mu_4} = \{z \in \mathbb R^4 \mid [0.5, 0, -\pi, -0.6]^T \leq z \leq [1, 1, \pi, 0.6]^T\}$, and $\mathbb S_{\mu_3} = \mathbb S_{\mu_{3, 1}} \cup \mathbb S_{\mu_{3, 2}} \cup \mathbb S_{\mu_{3, 3}}$, where
	$\mathbb S_{\mu_{3, 1}} = \{z \in \mathbb R^4 \mid [-2, -3, -\pi, -0.6]^T \leq z \leq [-1.3, 0, \pi, 0.6]^T\}$,
	$\mathbb S_{\mu_{3, 2}} = \{z \in \mathbb R^4 \mid [1.3, -1.5, -\pi, -0.6]^T \leq z \leq [2, 0, \pi, 0.6]^T\}$,
	$\mathbb S_{\mu_{3, 3}} = \{z \in \mathbb R^4 \mid [1.3, -3, -\pi, -0.6]^T \leq z \leq [2, -2, \pi, 0.6]^T\}$.
	
	\begin{figure}
		\centering
		\subfigure{ \includegraphics[width=0.35\textwidth]{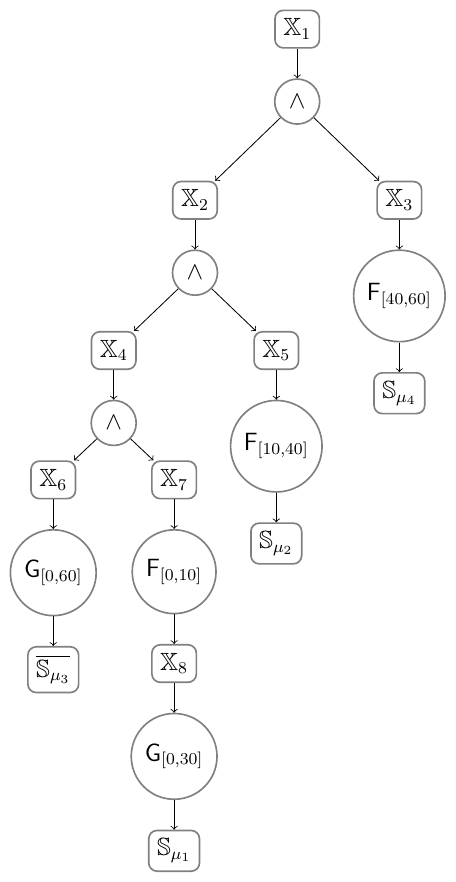}}
		\caption{\footnotesize The constructed tTLT $\mathcal{T}_{\varphi_{\rm parking}}$.}
		\label{Fig:parkingTLT}
	\end{figure}
	
	We let the full scenario be 60 seconds long and specify that the vehicle needs to enter the parking lot, park into the designated spot, and leave the parking lot within 10 seconds, 40 seconds, and 60 seconds, respectively. Then, this parking task can be encoded into the following STL formula:
	\begin{multline}
		\varphi_{\rm parking} = \mathsf{G}_{[0, 60]} \neg {\mu_3} \wedge \mathsf{F}_{[0, 10]}\mathsf{G}_{[0, 30]}{\mu_1} \wedge \mathsf{F}_{[10, 40]} {\mu_2} \\ \wedge \mathsf{F}_{[40,60]} {\mu_4}.
	\end{multline}
	First, we use Algorithm 1 to construct the corresponding tTLT $\mathcal{T}_{\varphi_{\rm parking}}$ (see Figure \ref{Fig:parkingTLT}), where the tube nodes $\mathbb{X}_i, i=1, \cdots, 8$ are computed in a bottom-up manner as in the previous example. Then, we implement the online control synthesis algorithm (Algorithm 5) on a SVEA vehicle using $\mathcal{T}_{\varphi_{\rm parking}}$. For choosing a control policy within the constraints of the synthesized control sets, we apply the same approach as described in Section IV.C of~\citet{jiang2020}.
	
	\begin{figure}
		\centering
		\includegraphics[width=0.45\textwidth]{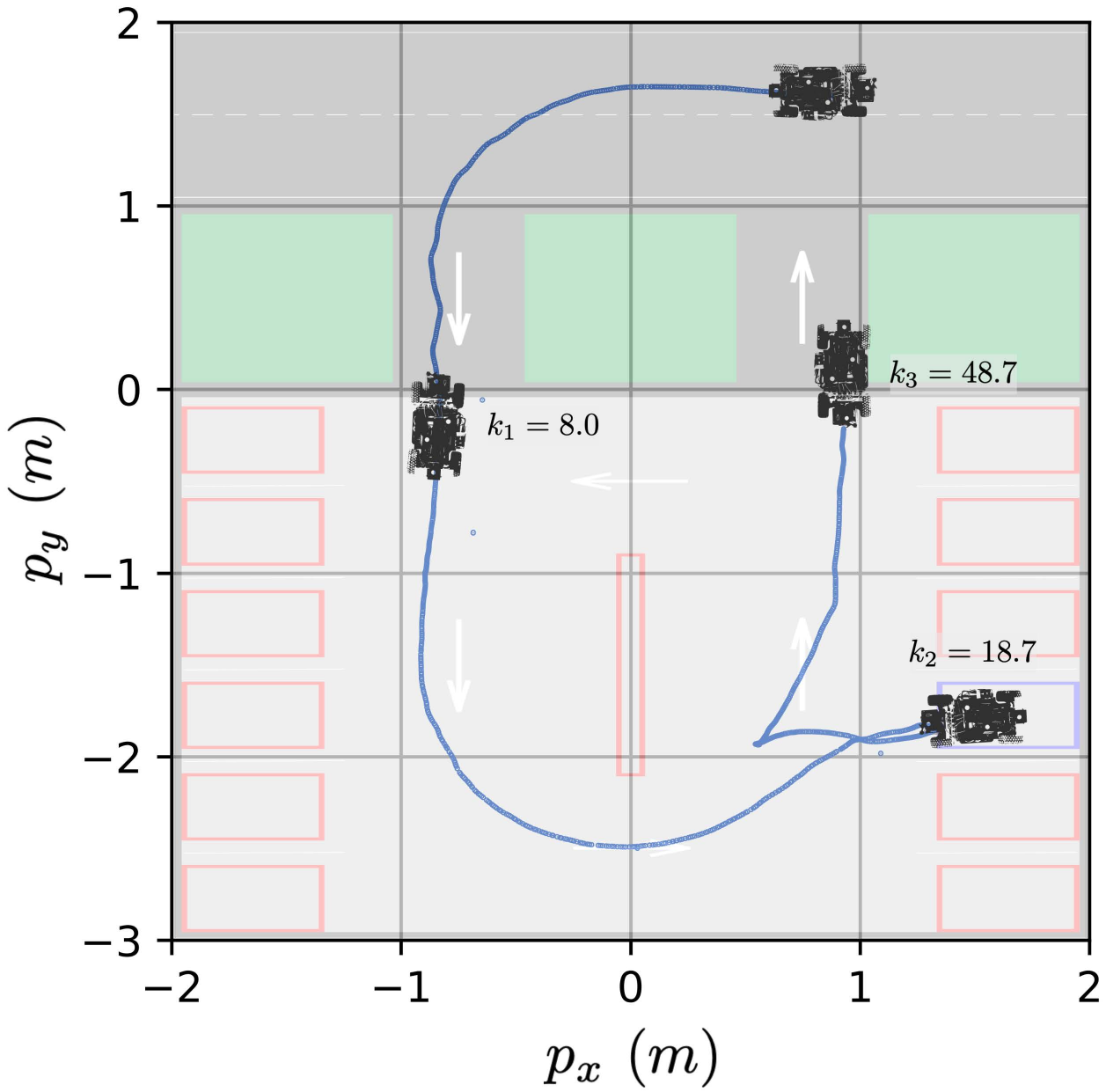}
		\caption{The position trajectory of a SVEA vehicle performing the parking task $\varphi_{\rm parking}$.}
		\label{Fig:xy_park}
	\end{figure}
	
\begin{figure*}
	\centering
	\includegraphics[width=0.85\textwidth]{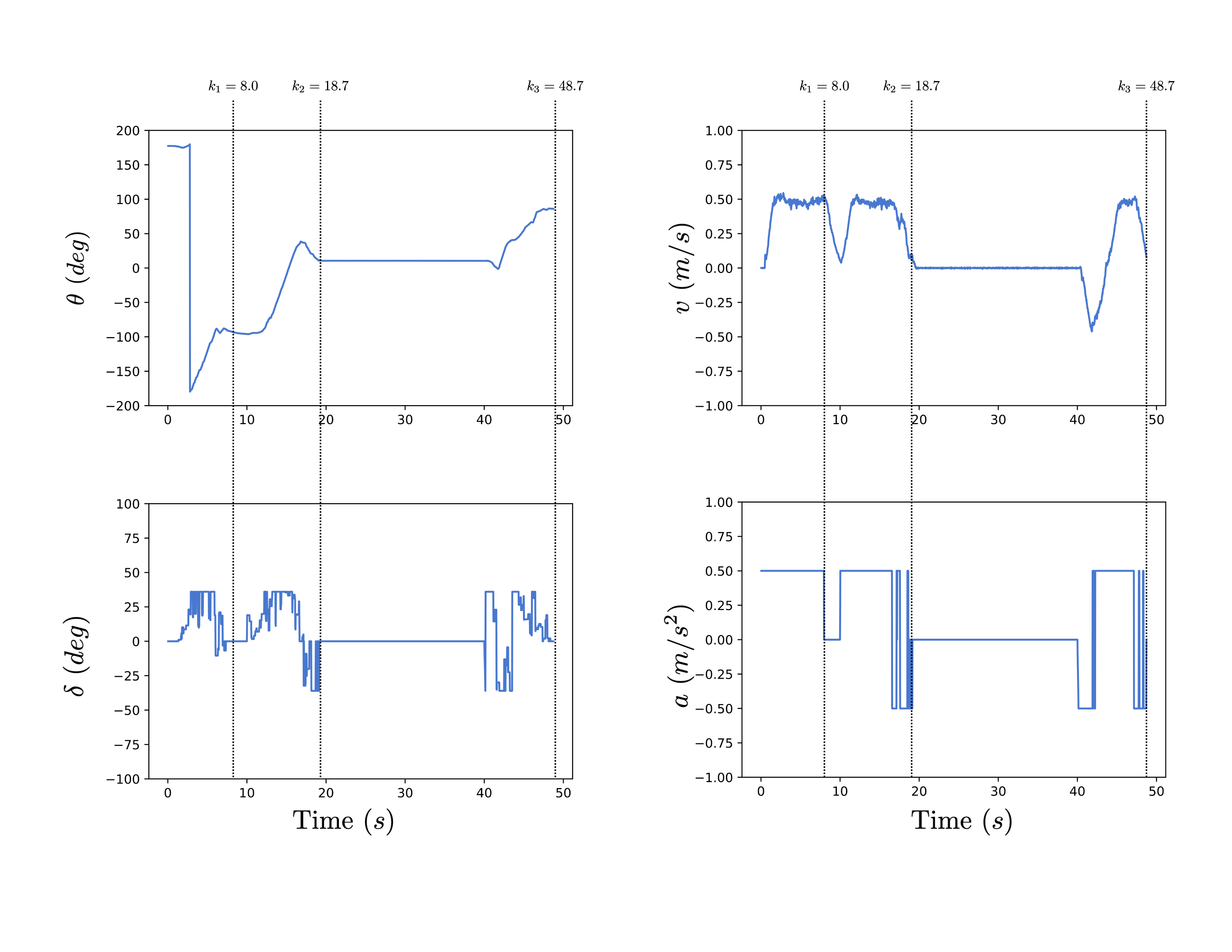}
	\caption{The velocity and heading trajectories in response to the acceleration and steering inputs throughout the parking task $\varphi_{\rm parking}$.}
	\label{fig:input_park}
\end{figure*}

	For our evaluation, we initialize the SVEA vehicle with the initial state of $x_0 = [1, 1.75, -\pi, 0]$. At this initial state, $\varphi_3$ is robustly satisfiable. Figure~\ref{Fig:xy_park} shows the position trajectory, where one can see that the specification is fulfilled. In Figure~\ref{fig:input_park}, we show the control input trajectories for acceleration and steering. We use $k_1$, $k_2$, $k_3$ to denote the minimal time instants that the automated vehicle reaches sets $\mathbb S_{\mu_1}$, $\mathbb S_{\mu_2}$, and $\mathbb S_{\mu_4}$. Using the synthesized controller, the SVEA vehicle realized $k_1 = 8.0$, $k_2 = 18.7$, and $k_3 = 48.7$, as illustrated in both Figures~\ref{Fig:xy_park} and~\ref{fig:input_park}, confirming the satisfaction of $\varphi_{\rm parking}$. For our evaluation, we initialize the SVEA vehicle with the initial state of $x_0 = [1, 1.75, -\pi, 0]$. At this initial state, $\varphi_3$ is robustly satisfiable. Figure~\ref{Fig:xy_park} shows the position trajectory, where one can see that the specification is fulfilled. In Figure~\ref{fig:input_park}, we show the control input trajectories for acceleration and steering. We use $k_1$, $k_2$, $k_3$ to denote the minimal time instants that the automated vehicle reaches sets $\mathbb S_{\mu_1}$, $\mathbb S_{\mu_2}$, and $\mathbb S_{\mu_4}$. Using the synthesized controller, the SVEA vehicle realized $k_1 = 8.0$, $k_2 = 18.7$, and $k = 48.7$, as illustrated in both Figures~\ref{Fig:xy_park} and~\ref{fig:input_park}, confirming the satisfaction of $\varphi_{\rm parking}$. 
	
Finally, we report the computation time of this example, which was run in Matlab R2022b with the Level Set Method Toolbox~\citep{mitchell2005toolbox}. We perform reachability analysis for constructing the tTLT offline on a Dell laptop with Ubuntu 20.04, Intel i7-4600U CPU 2.10GHz and 8.0 GB RAM, which takes 2371.81 seconds. We note that the offline computation time for constructing the tTLT can be significantly reduced by using the python implementation~\citep{bui2022}. Throughout the parking task, we perform the online control synthesis on an NVIDIA Jetson TX2 embedded computer onboard the SVEA vehicle. The average time step of the online control synthesis is 0.001 seconds. A video demonstration of this experiment can be found at \url{https://bit.ly/STLtTLT}.

\section{Conclusion}
A novel approach for the online control synthesis of uncertain discrete-time systems under STL specifications was proposed in this paper. Firstly, a real-time version
	of STL semantics and a notion of tTLT were introduced. Then the formal semantic connection between an STL
	formula and its corresponding tTLT was derived, \textit{i.e.}, a trajectory satisfying
	an tTLT also satisfies the corresponding STL formula.
	Finally, an online control synthesis algorithm was designed for the uncertain systems based on the connection between STL and tTLT. For the fragment of STL formulas under consideration, the soundness of the algorithm was proven. In the future, the control synthesis for multi-agent systems under local and/or global STL specifications is of interest.

\section*{Declaration of conflicting interests}

The author(s) declared no potential conflicts of interest with respect to the research, authorship, and/or publication of this article.

\section*{Funding}
The author(s) disclosed receipt of the following financial support for the research, authorship, and/or publication of this article: 
This work was supported by the Swedish Research Council (VR), the Swedish Foundation for Strategic Research (SSF), the Knut and Alice Wallenberg Foundation (KAW), and the ERC CoG LEAFHOUND.



\bibliographystyle{SageH}
\bibliography{reference}

\end{document}